# Study of vibrational kinetics of $CO_2$ and CO in $CO_2$-$O_2$ plasmas under non-equilibrium conditions

C. Fromentin[1], T. Silva[1], T. C. Dias[1], A.S. Morillo-Candas*[2], O. Biondo[3], O. Guaitella[2] and V. Guerra[1]

[1] Instituto de Plasmas e Fusão Nuclear, Instituto Superior Técnico, Universidade de Lisboa, Portugal

[2] Laboratoire de Physique des Plasmas (UMR 7648), CNRS, Univ. Paris Saclay, Sorbonne Université,

École Polytechnique, France

[3] Plasma Lab for Applications in Sustainability and Medicine – ANTwerp, Belgium

## Abstract

This work explores the effect of $O_2$ addition on $CO_2$ dissociation and on the vibrational kinetics of $CO_2$ and CO under various non-equilibrium plasma conditions. A self-consistent model, previously validated for pure $CO_2$ discharges, is further extended by adding the vibrational kinetics of CO, including electron impact excitation and de-excitation (e-V), vibration-to-translation relaxation (V-T) and vibration-to-vibration energy exchange (V-V) processes. The vibrational kinetics considered include levels up to $v = 10$ for CO and up to $v_1=2$ and $v_2=v_3=5$, respectively for the symmetric stretch, bending and asymmetric stretch modes of $CO_2$, and accounts for e-V, V-T in collisions between CO, $CO_2$ and $O_2$ molecules and O atoms and V-V processes involving all possible transfers involving $CO_2$ and CO molecules. The kinetic scheme is validated by comparing the model predictions with recent experimental data measured in a DC glow discharge, operating at pressures in the range 0.4 - 5 Torr (53.33 - 666.66 Pa). The experimental results show a lower vibrational temperature of the different modes of $CO_2$ and a decreased dissociation fraction of $CO_2$ when $O_2$ is added to the plasma but an increase of the vibrational temperature of CO. On the one hand, the simulations suggest that the former effect is the result of the stronger V-T energy-transfer collisions with O atoms which leads to an increase of the relaxation of the $CO_2$ vibrational modes; On the other hand, the back reactions with $O_2$ contribute to the lower $CO_2$ dissociation fraction with increased $O_2$ content in the mixture.

Keywords: vibrational kinetics, DC glow discharge, reaction mechanism, $CO_2$ conversion, low-temperature plasma, model validation

## I. Introduction

The growing concentration of greenhouse gases in the atmosphere coming from anthropogenic activities [1], and the resulting climate change are a great concern of our century. As $CO_2$ has the most important contribution to global warming [2], it is necessary to focus on reducing its concentration in the atmosphere via Carbon Capture and Utilization (CCU) for chemical synthesis and fuel production [3], ideally using electricity from renewable energy sources, and $CO_2$ from industrial emissions as a feedstock. However, $CO_2$ is, thermodynamically, a very stable molecule so the conversion of $CO_2$ is limited by the initial dissociation step ($CO_2 \rightarrow CO + O$). The conventional heating of the gas can be used in principle to split $CO_2$ molecules. However, the high energy cost of this method, among other limitations, make technologies like non-thermal plasmas (NTP) very attractive for $CO_2$ conversion [4-10]. NTPs are characterized by non-equilibrium conditions, where high energy electrons and cold heavy species concur, which are ideal for the breaking of chemical bonds while they can be operated at room temperature and atmospheric pressure.

$CO_2$-containing discharges are intensively studied nowadays, both in terms of experimental work and modelling [5, 9-24] to bring new insights into the kinetics of $CO_2$ dissociation by different pathways. The CO2 dissociation by direct electron impact requires at least 7eV and produces CO and O in an

*Current affiliation: Paul Scherrer Institut, CH-5232 Villigen PSI, Switzerland

electronically excited state [25]. However, by taking advantage of non-equilibrium plasma processes, only 5.5 eV may be required to obtain the products in the ground state via stepwise vibrational excitation of $CO_2$ by anharmonic VV up pumping to the dissociation limit [26].

Following the measurements of Klarenaar *et al.* in a pulsed $CO_2$ glow discharge [16], T. Silva *et al.* studied the complex kinetics of the relaxation of vibrationally excited $CO_2$ levels during the afterglow validating a set of V-T and V-V energy transfer processes and the corresponding rate coefficients [17]. Moreover, in [18] the investigation focuses as well on the active discharge, by extending the model with the inclusion of electron impact processes for vibrational excitation and de-excitation (e-V) [17-19]. In parallel, A.F. Silva *et al.* established a reaction mechanism (i.e., a set of reactions and rate coefficients validated against benchmark experiments) for 'vibrationally cold $CO_2$ plasmas', considering the $CO_2$ dissociation products, validated by comparing simulation results with experimental data measured in continuous $CO_2$ glow discharges where dissociation cannot be neglected [14]. Other modelling research works focused on the electron-neutral scattering cross sections for $CO_2$ [27] and CO [28], the electron-impact dissociation cross sections of $CO_2$ [15], the dynamics of gas heating in the afterglow of pulsed $CO_2$ and $CO_2$–$N_2$ glow discharges at low pressure further validating the V-V and V-T mechanisms and rate coefficients [21], the $CO_2$ dissociation under Martian environment for oxygen production [11, 29, 30] and the role of electronically excited metastable states in $CO_2$ dissociation and recombination [31]. More information about these works can be found in [22] where recent advances in non-equilibrium $CO_2$ plasma kinetics are reviewed.

Herein we extend the study of the coupled electron, vibrational and chemical kinetics developed in [11, 14, 17-19] with the addition and validation of the CO vibrational kinetics, by including 10 vibrational levels of CO, and an accurate description of the vibrational kinetics involving the dissociation products, namely CO, $O_2$, O as was initiated in [11]. This constitutes a major improvement regarding our previous simulations for $CO_2$ plasmas and is relevant as CO is a product of the $CO_2$ dissociation and therefore always present in $CO_2$ gas discharges. In parallel, we address the study of $CO_2$-$O_2$ mixtures. Indeed, investigating the influence of $O_2$ on the $CO_2$ dissociation is relevant as $O_2$ is an impurity often present in industrial emissions [32]. In addition, $O_2$ is also one of the main by-products of the dissociation of $CO_2$, formed from the recombination of O atoms. The admixture of $O_2$ has a detrimental impact on $CO_2$ decomposition, as shown experimentally in [12], as it leads to a decrease of the $CO_2$ dissociation fraction via the enhancement of the reverse reaction ($CO(a^3\prod_r) + O_2 \rightarrow CO_2 + O$), producing back $CO_2$ from electronically excited CO, $CO(a^3\prod_r)$, in collisions with $O_2$ [13]. Besides, the presence of oxygen in the discharge influences greatly the vibrational kinetics of $CO_2$ and CO mostly via the quenching with O atoms [33]. Finally, by varying the $O_2$ content in $CO_2$-$O_2$ mixtures we enlarge the parameter space and can have a thorough validation of the model and gain a deeper understanding of the kinetics of $CO_2$ plasmas.

To establish a reaction mechanism for vibrationally excited $CO_2$ and $CO_2$-$O_2$ plasmas a DC glow discharge (plasma sustained by high voltages inside a pair of electrodes) is used as it generates a stable (axially) homogeneous plasma (in the positive column) and is accessible to different diagnostics and therefore optimal for model validation. The $CO_2$ and CO densities and its vibrational kinetics are diagnosed by FTIR spectroscopy, and actinometry is used to determine the O atom density and O loss frequency.

The paper is structured as follows. Section 2 provides information about the experimental setup and the diagnostics used. In section 3 the model is described, and the kinetic scheme used to study the $CO_2$ discharge is specified. Moreover, in this section we also detail some rate coefficients for electron impact reactions, vibration-translation and vibration-vibration exchanges involving $CO_2$ and CO. The comparison between the experiments and the simulations is presented and discussed in section 4 to gain further insight into the underlying kinetics. Finally, section 5 summarizes the main findings of this work.

## II. Experiment

The experimental setup used to obtain the data described in this work consists of a DC glow discharge ignited in a cylindrical Pyrex tube of 1 cm radius. Two different reactor lengths were used, 67 cm for actinometry and 23 cm for in situ Fourier Transform Infrared (FTIR) spectroscopy experiments with the electrodes positioned respectively 53 or 17 cm apart, respectively, depending on the tube length and opposite to the gas in- and outlet. Two discharge currents were used, 20 and 40 mA and the pressure varied between 0.4 and 5 Torr, using a scroll pump (Edwards XDS-35), and a pressure gauge (Pfeiffer CMR263) with feedback to an automated pressure regulating valve (Pfeiffer EVR116) and controller (Pfeiffer RVC300). The reactor is connected in series with a 40 kΩ resistor to a DC power supply. The gas flows are controlled using mass flow controllers (Bronkhorst F-201CV). A total gas flow of 7.4 sccm is used as the reference condition in the present experiments as previously employed in [11, 16, 20, 34, 35]. The experimental set-up and measurement techniques (actinometry and FTIR spectroscopy) are presented and described in detail in [16, 34].

The $CO_2$ and CO vibrational and rotational temperatures and dissociation fraction are obtained using in situ Fourier transform infrared (FTIR) spectroscopy as an outcome of the fitting of the measured IR spectra containing several lines of CO and $CO_2$ vibrational transitions, as described by Klarenaar *et al.* in [16, 35], in a 23 cm long reactor. The rotational temperature ($T_{rot}$) can be assumed to be in equilibrium with the gas temperature ($T_g$) [34] and is used as an input parameter for our model. The quantities obtained by FTIR spectroscopy are approximately an average over the radius of the reactor since the FTIR beam fills most of the discharge tube and it is assumed that the rotational and vibrational temperatures are uniform along the length of the reactor. The sensitivity of the fitted transmittance to the different temperatures can give an indication of their error, which was estimated to be 30 K and 27 K for $T_{rot}$ and $T_{1,2}$ respectively, 67 K for $T_3$ and 357 K for $T_{CO}$ at 5 Torr, 50 mA and in pure $CO_2$ [16].

The average electric field in the plasma bulk is estimated by measuring the voltage drop in the positive column, considered homogeneous, between two tungsten probes, at the floating potential, pointing radially inside the reactor.

The experimental characterization of the discharge comprises the determination of O atom densities and O loss frequencies by actinometry measured in a 67 cm length tube [34]. The measured loss frequencies, $\nu_{loss}$, can be the result of both surface loss processes and/or gas phase reactions [34]:

$$\nu_{loss} = \frac{v_{th} \cdot \gamma_O}{2R} + L_{gp}, \tag{1}$$

where $L_{gp}$ represents the contribution of the gas phase losses, $\gamma_O$ is the O atom surface loss probability, $v_{th}$ is the thermal velocity of the O atoms and R is the radius of the discharge tube. In the present conditions the contribution of gas phase losses can be discarded [34] and expression (1) becomes:

$$\gamma_O = \frac{2R \cdot \nu_{loss}}{v_{th}}, \tag{2}$$

where

$$v_{th} = \sqrt{\frac{8 \cdot k_B \cdot T_g}{\pi \cdot m}}, \tag{3}$$

$k_B$ is the Boltzmann constant, m the mass and $T_g$ the gas temperature. The error on the loss frequency (from which is calculated the recombination probability, c.f. equation (2)), related to the reproducibility of the experiment, is of the order of 15%. The O densities measured using actinometry rely on many rate coefficients and on the choice of electron impact excitation cross sections and are given with a minimum error of 30% [34].

In this work we will discuss three data sets. In the first one we have varied the $CO_2$-$O_2$ gas mixture (Air Liquide Alphagaz 1 for $CO_2$ a Alphagaz 2 for $O_2$). The FTIR experiments were done in the 23 cm reactor and the error bars on $T_{CO}$ being so large when $O_2$ is added to the mixture we only provide the experimental data for the pure $CO_2$ case. The large uncertainty on the CO temperatures measured by FTIR is due to the low signal to noise ratio for the CO band particularly at low pressures and currents and high $O_2$ content, corresponding respectively to low particle density and low CO concentrations. The relative fitting error on the dissociation fraction was calculated and is lower than 2% for all the conditions measured. The O atom surface loss probabilities, $\gamma_O$, used in this work for the simulations of the $CO_2$-$O_2$ mixture can be found in Table 1. They were measured in the 67 cm reactor just as the O/N. It is important to note that the loss frequencies were obtained for a wall temperature $T_{wall}$ of the Pyrex tube of 25 ºC whereas the temperatures obtained by FTIR spectroscopy were obtained for $T_{wall}$=50 ºC. The effect was verified to be around 30 K for $T_3$ and 20 K for $T_{1,2}$ and therefore considered negligible as it lays within the experimental error.

*Table 1: O atom surface loss probabilities, $\gamma_O$, for different $CO_2$-$O_2$ mixtures, pressures, and discharge currents, calculated from the experimental loss frequency measurements. *extrapolated values.*

|  |  | $CO_2$ initial fraction | | | |
| --- | --- | --- | --- | --- | --- |
|  |  | 0.25 | 0.5 | 0.75 | 1 |
| Pressure (Torr) | Current (mA) | O loss probability | | | |
| 1 | 20 | 0.000454458 | 0.000386444 | 0.000282906 | 0.000212388 |
| 2 | 20 | 0.000531294 | 0.000445574 | 0.000290678 | 0.000201258 |
| 5 | 20 | 0.000697701* | 0.000662973 | 0.000427483 | 0.000353317 |
| 1 | 40 | 0.000739642 | 0.000616286 | 0.000468980 | 0.000284431 |
| 2 | 40 | 0.000917035 | 0.000645062 | 0.000524091 | 0.000357731 |
| 5 | 40 | 0.001163260* | 0.000925654 | 0.000650424 | 0.000466082 |

The $CO_2$ dissociation fraction obtained in the short tube can be extrapolated for the longer tube (67cm), where the O atom measurements were taken. This was possible by measuring the dissociation fraction and vibrational and rotational temperatures as a function of the residence time since downstream measurements using both long and short tubes confirmed that the $CO_2$ dissociation fraction is the same in both reactors, for a given residence time, in pure $CO_2$ [34]. Moreover, the temperatures for the same pressure and current are assumed to be the same in both reactors due to the fast timescales of temperature evolution in comparison with the residence times in the experimental conditions.

Another data set presented in [33] is also analysed in section IV.4. It consists of a Pyrex tube (23 cm length) covered with micro-structured silica fibers enhancing the O recombination at the walls. The vibrational and rotational temperatures of $CO_2$ and CO were measured with in situ FTIR spectroscopy. For this study we adopt the surface O loss probabilities, $\gamma_O$, from [34]. Finally, we recall a third dataset, containing the vibrational and rotational temperatures of $CO_2$ and CO, O/N, E/N and CO/N and surface O loss probabilities, $\gamma_O$, for the pure $CO_2$ case since it contains data for an extended pressure range and can be found in [34].

## III. Model

### 1) General formulation

The self-consistent global model used in this work couples the homogeneous two-terms approximation Boltzmann equation for the electrons to a set of zero-dimensional (spatially averaged) rate balance equations describing the creation and destruction of the neutral and charged heavy species considered. The simulations are performed with the Lisbon Kinetics (LoKI) [36,37] numerical code, composed of two modules:

LoKI-B: solves the time and space independent electron Boltzmann equation within the two-terms approximation, for non-magnetised non-equilibrium low-temperature plasmas (LTPs) excited by DC/HF electric fields for different gases or gas mixtures and provides the electron energy distribution function (EEDF), electron transport parameters and electron impact rate coefficients;

LoKI-C: solves a system of zero-dimensional rate balance equations for the heavy species.

The electron, chemical and vibrational kinetics are coupled into a self-consistent scheme for which the reduced electric field, E/N, corresponds to steady-state conditions where the total rate of production of electrons in ionization events must compensate exactly their total loss rate due to ambipolar diffusion to the wall and electron-ion recombination, while satisfying the quasi-neutrality condition.

The diffusion scheme adopted to describe the charged-particles losses is the ambipolar diffusion to the reactor walls. For the heavy species, including the vibrationally excited species, we use the Chantry model [38] to obtain the loss rate of a particular species interacting with the wall due to the combined effect of transport (with a diffusion coefficient) and the reaction at the wall (with a certain wall recombination/deactivation probability γ) [20, 39]. The renewal of the gas in the reactor influences the densities of the species in the plasma and was thus included in the model. The rate coefficient for the inlet and outlet flow of species is calculated assuming conservation of atoms in the gas/plasma mixture as described in [14]: new $CO_2/O_2$ particles enter the reactor while the species produced in the plasma exit at the outlet.

The input parameters of the model are the gas pressure (P), discharge current (I) and the initial gas mixture and corresponding gas flows controlled during the experiment (see section II), as well as the dimensions the experimental reactor. The loss probability of O atoms at the wall, $\gamma_O$, is also included as input parameter and deduced from the experimental determination of O-atom loss frequencies. Additionally, in the present simulations the gas temperature is also given as an input parameter since its value is available from experiment and our purpose is not to focus on the gas heating mechanisms but rather on the plasma chemistry. However, the gas thermal balance equation can be incorporated in the current formulation of the model as already done in [21, 40, 41] for the study of gas heating mechanisms. The average electron density was calculated based on the discharge current and the electron drift velocity obtained from the Boltzmann equation solution.

### 2) Kinetic scheme

A kinetic description of both electrons and heavy species is needed to accurately describe the plasma under study. For the electron kinetics we use a complete and consistent set of cross sections from the IST-Lisbon group available on the open-access website LXCat [42] and described in [27] (for $CO_2$), [43, 44] (for O and $O_2$) and [28] (for CO). It is worth noting that in the Boltzmann solver the superelastic electronic collisions with the different rotational, vibrational and electronic states of these molecules are taken into account.

Note that the Polak and Slovetsky's total cross sections for electron-impact dissociation of $CO_2$ [25] are not part of the complete and consistent $CO_2$ cross section set and, accordingly, are not used to obtain the EEDF, but are integrated with the calculated EEDF to obtain the corresponding rate coefficient as suggested in [15]. Indeed Morillo-Candas *et al.* [15] validated the electron impact $CO_2$ dissociation cross sections, in the range of reduced electric fields 40-110 Td using two complementary methods: through the comparison of the measured rate coefficients in a large range of reduced electric fields with those derived from cross sections, available in literature; and through the comparison of the experimental time evolution of the dissociation fraction with the simulations of a 0D model and thus recommend the use of these cross sections for the calculation of the $CO_2$ electron impact dissociation rate under those discharge conditions.

The complex plasma chemistry used in this work is based on previous publications dealing with $CO_2$ vibrations [17-19], kinetic mechanisms in $O_2$ plasmas [45], and plasma chemistry in vibrationally cold $CO_2$ [14] and includes the following species: ground-state and electronically excited CO, $CO_2$ and $O_2$ molecules $CO(X^1\Sigma^+)$, $CO(a^3\Pi_r)$, $CO_2(X^1\Sigma^+_g)$, $O_2(X^3\Sigma_g^-)$, $O_2(a^1\Delta_g)$, $O_2(b^1\Sigma_g^+)$; ground-state and electronically excited oxygen atoms, $O(^3P)$, $O(^1D)$, ground-state ozone and vibrationally excited ozone, $O_3$, $O_3^*$; and positive and negative ions, $O^+$, $O_2^+$, $O^-$, $CO_2^+$, $CO^+$. For $O_3^*$ we consider a single effective vibrationally excited state [46]. For the kinetics of oxygen, the set proposed in [45] is adopted without modifications except for the exclusion of vibrational states in the heavy species chemistry and the use of the measured loss frequency of the ground state of atomic oxygen. We further use the chemistry set proposed by Silva *et al.* in [14] for vibrationally-cold low-pressure $CO_2$ plasmas, to which we added the three-body reactions:

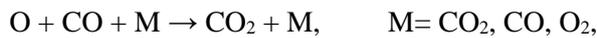

O + CO + M → $CO_2$ + M,     M= $CO_2$, CO, $O_2$,

with the rate coefficients taken from [47].

In the present conditions of pressures and temperatures, the three-body reactions play a negligible role [14] and the dominant "back reaction" should be a 2-body mechanism [48]. However, three-body processes could be relevant to properly describe the recombination of $CO_2$ at high pressure conditions, of interest for plasma reforming for instance; therefore, we include them in the current formulation of the model for completeness.

### 3) Vibrational kinetics

The present work studies the state-to-state kinetics of the first 72 low-lying levels of $CO_2$ plasma during the active discharge, corresponding to $CO_2$ ($v_1^{max}=2$, $v_2^{max}=v_3^{max}=5$), where $v_1$, $v_2$ and $v_3$ are quantum numbers of the symmetric stretching, bending, and asymmetric stretching vibrational modes, respectively, with energies up to about 2 eV and the first 10 levels of CO with energy up to about 2.5eV [17] and a population of around $10^{-6}$ for $v$=10. The vibrational kinetics of $O_2$ are only considered in the Boltzmann solver to obtain the EEDF, assuming a Boltzmann distribution at the gas temperature, while the vibrational kinetics of $CO_2$ and CO are included as well in the rate balance equations, as it improves the calculation of the EEDF and was taken into account when the consistent set of cross sections for $O_2$ was determined. Unlike Annušová et al. [45], we do not include vibrationally excited $O_2$ molecules for the chemistry part as it does not influence the simulation results (densities, E/N etc) for our conditions but would significantly increase the computation time. Indeed, the vibrational distribution function of $O_2$ shows a steep decrease already at low vibrational levels (reaching $\sim 10^{-6}$ at $v$=2).

The density of the different vibrationally excited levels is governed by the rate of creation and loss by electron impact, vibrational-translational and vibrational-vibrational exchanges and chemical reactions and processes like dissociation or ionization [18]. Due to the lack of data, the dissociation cross sections via electron impact from vibrationally excited states are considered with a threshold shift, while keeping the same amplitude as for dissociation from the ground-state [11]. The same procedure is used for ionization from vibrationally excited CO and $CO_2$ molecules.

Following the approach in [17-19], the $CO_2$ vibrational levels are described by four quantum numbers using the notation $CO_2(v_1v_2^{l_2}v_3f)$, also known as Herzberg's form [49]. In the present work the $CO_2$ vibrational levels under Fermi resonance are considered as one single effective level. The Fermi resonance refers to an accidental energy degeneracy between certain vibrational modes. In the case of $CO_2$, the modes $v_1$ and $2*v_2$ have very close vibrational energies, resulting in a coupling between the $CO_2(v_1v_2^{l_2}v_3)$ and $CO_2((v_1-1)(v_2+2)^{l_2}v_3)$ levels to form new states that are assumed to be in local equilibrium. All the vibrational levels coupled together have the same orbital quantum number $l_2$ (projection of the angular momentum of bending vibrations) since those with different $l_2$ cannot perturb each other. The ranking number $f$ is always equal to $v_1+1$ and indicates how many individual levels are accounted for in the effective level.

The energies of the individual levels are calculated according to the anharmonic oscillator approximation and are based on [50] using the spectroscopic constants from the same reference. The calculated values were compared with experimental spectroscopic data available in [51] and show a good agreement.

The vibrational energy of the effective level is determined through the average of the vibrational energies of all the individual levels in the effective level and we assume that the average energy of unperturbed levels is the same as the average energy of the levels perturbed by the Fermi resonance coupling. The statistical weight is determined through the sum of the statistical weights of the individual states.

The CO vibrational energy levels can be calculated by the formula [52]:

$$\frac{E_{co}}{hc} = \omega_e(v + 0.5) - \omega_e x_e(v + 0.5)^2 \qquad (4)$$

where $v$ is the vibrational quantum number, $\omega_e$ is the vibrational frequency and $x_e$ is the non-dimensional anharmonicity. We use the values $\omega_e$ =2169.81 cm$^{-1}$ and $\omega_e x_e$ =13.29 cm$^{-1}$, obtained from the NIST Chemistry WebBook [53]. Only the first 10 vibrational levels of CO, up to the energy of 2.51 eV, are included in the model as it is already higher than the energy of the highest $CO_2$ vibrational level included. Moreover, they are enough to calculate accurately the vibrational temperatures of CO and $CO_2$ in the present conditions (c.f. section IV), as the relative population of $v$=10 is always smaller than $5 \cdot 10^{-6}$.

**e-V**

The cross sections for the e-V reactions included in our model are obtained from a direct deconvolution of the available lumped cross sections according to the statistical weights of the various levels, as reported by Grofulović *et al.* [18, 27]. The excitation cross-sections of most of the vibrational levels considered are unknown, but they can be generated using the Fridman approximation [26] if the cross-section for the excitation from the ground state to the first excited state is known. The rate coefficient $C_{ij}$ of the excitation from $CO_2(000i1)$ to $CO_2(000j1)$ is given by:

$$C_{ij} = \frac{\exp[-\alpha_f(j - i - 1)]}{1 + \beta i} C_{01}, \qquad (5)$$

with $C_{01}$ the rate coefficient for the excitation from the ground state to the first excited state. The Fridman approximation scales the magnitude of the rate coefficients according to two parameters, $\alpha_f$ and $\beta$. Due to lack of data for excitation from $CO_2(00011)$ to higher levels, we have no information of the $\beta$ value and for simplicity we use $\beta = 0$, i.e. the cross section $\sigma_{12}$ has the same magnitude as $\sigma_{01}$, and $\alpha_f = 0.5$ [26]. To avoid an overpopulation of the vibrational distribution associated to the higher levels ($v_3 \geq 3$) of the asymmetric mode of $CO_2$, we investigated the possibility of setting $\alpha_f$ to 3 in expression (5) for the transitions e + $CO_2(00001) \leftrightarrow$ e $CO_2(000v_31)$ with $v_3$=3, 4 and 5. Indeed, we saw unrealistically high populations of $v_3$=3, 4, 5 when using $\alpha_f$ =0,5 for all transitions while the calculated Vibrational Distribution Functions (VDFs) were in good agreement with experimental ones, obtained using the FTIR setup described in the experimental section, when using $\alpha_f = 3$ for $v_3 \geq 3$. Note that this overpopulation is also strongly dependent on the scaling law used to describe the V-V rates. Using the cross sections from Laporta et al. [54] for the asymmetric mode $v_3$, calculated only for resonant transitions, does not improve the results because non resonant contributions also play an important role as stated in [54]. Setting $\alpha_f$ to 3 for higher levels improves the shape of the VDF and leads to a good agreement between the vibrational temperatures from the model and from the experiment but it means that the corresponding rate coefficients become very small. An investigation of the influence of the eV processes on the VDF will be carried out in a future work but for this study we use $\alpha_f = 3$ for the transitions e + $CO_2$(00001) ↔ e + $CO_2$(000$v_3$1) with $v_3$=3,4 and 5. Note that in Table 5 of [18], the Fridman approximation is applied as (5) and only for the reactions number 7 to 10. For the other reactions, the cross sections are obtained according to the description in the dedicated column from the original cross sections available in [27]. The rate coefficient of the reverse processes is calculated from the principle of the detailed balance by multiplying the coefficients of the direct processes times the ratio of the statistical weights of the final and initial states and by the Boltzman factor $e^{-\Delta E/k_B T_e}$, where $T_e$ is the electron temperature [55].

For the electron impact excitation of the CO vibrations, we have adopted the cross sections from [28] for the vibrational excitation and de-excitation which are largely based on resonant excitation data from Laporta and co-workers [56], and where contributions from non-resonant collisions for the transition e + CO($v$=0) ↔ e + CO($v$=1), taken from [57], are also included.

**Vibrational quenching on the wall**

An important phenomenon in the vibrational kinetics of $CO_2$ in our experimental conditions is the deactivation of vibrationally excited $CO_2$ and CO molecules through collisions on the wall. Deactivation of vibrationally excited states on the walls is shown to have a significant influence on the vibrational characteristic temperatures especially for pressures below 1 Torr [20].

Following [20] and due to the lack of experimental values, we set the same value of deactivation probability, $\gamma_v$, for any mode of $CO_2$ being deactivated to the ground state, i.e. $\gamma_v(CO_2(v>0)) = 0.2$ for a Pyrex surface (average value from table 1 of [58]). As it is assumed for $CO_2$, we use a constant value of $4 \cdot 10^{-2}$ for the deactivation probability for all levels of CO [58], which is significantly lower than for $CO_2$, as measurements showed that the probabilities of heterogeneous relaxation of CO do not depend on the value of $v$, at least for $v$ = 1, 2 and 3 [59]. A certain dependence on the vibrational level could be expected, similar to the case of $N_2$, where a linear dependence of $\gamma_v$ with the vibrational level is reported [60, 61]. Moreover, as opposed to $CO_2$, we consider single-quantum transitions as done for $N_2$ and $O_2$ in [41] where only one vibrational quantum is lost upon collision with the wall. No significant difference in the vibrational kinetics is expected for vibrational levels below $v$=10 between the assumption of single and multi-quanta relaxation. However, for higher vibrational levels, the best agreement of modelled and experimental CO VDFs was achieved for the 'multi-quanta' mechanism (loss of all vibrational quanta upon collision) [62].

**V-V and V-T processes**

One of the problems arising in the development of a state-to-state $CO_2$ model is the scarcity of data on the rate coefficients of different kinds of vibrational energy transitions within and between modes. For diatomic molecules, there are two main mechanisms of vibrational relaxation, namely, V-V exchanges of vibrational quanta and V-T transitions of vibrational energy to translation. However, since $CO_2$ is a polyatomic molecule, it has multiple vibrational modes and several additional relaxation channels like the inter-mode exchanges. Studying the vibrational kinetics of $CO_2$ thus requires a larger amount of data than for diatomic molecules.

Most of the data used for the V–T and V–V rate coefficients in our model are taken from the report of Blauer and Nickerson [63] regrouping experimental results and theoretical studies for the most important deactivation channels. This work provides rate coefficients (based on either experimental values or theoretically calculated results) for the first fourteen vibrational levels of $CO_2$ ($v_1^{max}$ = 2, $v_2^{max}$ = 5, $v_3^{max}$ =1). The authors have adapted the well-known SSH theory to the case of $CO_2$ vibrational energy transfers by considering the presence of Fermi resonance. Therefore, this report offers many rate coefficients for transitions involving changes for the $v_1$ or $v_2$ quantum numbers, while $v_3$ remains constant. Unfortunately, this is not sufficient for the description of transitions involving higher $v_3$ vibrational quantum number. For the missing reactions that cannot be found in literature, we determine the rate coefficients based on either SSH (Schwartz, Slawsky, Herzfeld) theory [64] or Sharma–Brau scaling [65] accounting for short-range contributions or describing transitions dominated by long-range interactions, respectively. Unlike the e-V processes described in the previous subsection, the V-V and V-T rates are derived only for single quantum exchanges and the complete set of V-T and V-V reactions for $CO_2$ can be found in Silva *et al.* [17]. On the one hand, the databases for multi-quanta exchanges are too scarce and there is no available experimental data for comparison; on the other hand, the SSH and SB theories predict a null rate coefficient for these exchanges, while the forced harmonic oscillator (FHO) calculations from [66] confirm that for the low gas temperatures pertinent to this study multi-quanta transitions can be safely disregarded. More information on these processes can be found in a recent study on the vibrational kinetics of $CO_2$ [40].

All the processes included in the model are listed here and are detailed in the following sections:

$VT_{CO2\text{-}CO2}$ : $CO_2(v_1 v_2^{l2} v_3 f) + CO_2 \leftrightarrow CO_2(v'_1 v'_2^{l2} v'_3 f) + CO_2$

$VT_{CO2\text{-}CO}$ : $CO_2(v_1 v_2^{l2} v_3 f) + CO \leftrightarrow CO_2(v'_1 v'_2^{l2} v'_3 f) + CO$

$VT_{CO2\text{-}O2}$ : $CO_2(v_1 v_2^{l2} v_3 f) + O_2 \leftrightarrow CO_2(v'_1 v'_2^{l2} v'_3 f) + O_2$

$VV_{CO2\text{-}CO2}$ : $CO_2(v_1 v_2^{l2} v_3 f) + CO_2(v_1 v_2^{l2} v_3 f) \leftrightarrow CO_2(v'_1 v'_2^{l2} v'_3 f) + CO_2(v'_1 v'_2^{l2} v'_3 f)$

$VV(SB)_{CO2\text{-}CO2}$ : $CO_2(v_1 v_2^{l2} v_3 f) + CO_2(v_1 v_2^{l2} v_3 f) \leftrightarrow CO_2(v_1 v_2^{l2} (v_3+1) f) + CO_2(v_1 v_2^{l2} (v_3-1) f)$

$VT_{3\text{-}O}$ : $CO_2(v_1 v_2^{l2} v_3 f) + O \leftrightarrow CO_2(v_1 (v_2=2,3,4)^{l2} (v_3-1) f) + O$

$VT_{2\text{-}O}$ : $CO_2(v_1 v_2^{l2} v_3 f) + O \rightarrow CO_2(v_1 (v_2-1)^{l2-1} v_3 f) + O$

$VV_{3\text{-}CO}$ : $CO_2(v_1 v_2^{l2} (v_3+1) f) + CO(w) \leftrightarrow CO_2(v_1 v_2^{l2} v_3 f) + CO(w+1)$

$VV_{1,2\text{-}CO}$ : $CO_2((v_1+1)(v_2+1)^{l2+1} v_3 f) + CO(w) \leftrightarrow CO_2(v_1 v_2^{l2} v_3 f) + CO(w+1)$

$VV_{CO\text{-}CO}$ : $CO(v) + CO(w-1) \leftrightarrow CO(v-1) + CO(w)$

$VT_{CO\text{-}CO}$ : $CO(v) + CO \leftrightarrow CO(v-1) + CO$

$VT_{CO\text{-}CO2}$ : $CO(v) + CO_2 \leftrightarrow CO(v-1) + CO_2$

$VT_{CO\text{-}O2}$ : $CO(v) + O_2 \leftrightarrow CO(v-1) + O_2$

$VT_{CO\text{-}O}$ : $CO(v) + O \leftrightarrow CO(v-1) + O$

*$CO_2$- $CO_2$ V-V / V-T*
There are roughly 350 V-T and 600 V-V direct processes to describe the kinetics of the 72 $CO_2$ vibrational states considered. The various coefficients are fitted through the following exponential expression [63]:

$$k(cm^3 s^{-1}) = 1{,}66 \cdot 10^{-24} \cdot \exp(a + b \cdot T^{-1/3} + c \cdot T^{-2/3}) \qquad (6)$$

where a, b and c are the fitting constants. The rate coefficients for the inverse reactions are calculated by the principle of detailed balance [55].

The transitions involving higher vibrational levels are scaled with the SSH theory, except for the nearly resonant collisional up-pumping process along the asymmetric stretching mode, given by:

$CO_2$ ($00^0 v_3 1$) + $CO_2$ ($00^0$ $v_3 1$) ↔ $CO_2$ ($00^0$ ($v_3$-1)1) + $CO_2$ ($00^0$ ($v_3$+1)1)

Indeed, the SB theory (based on long-range forces) was used (instead of SSH) to obtain an empirical formula for the rate constants as a function of the gas temperature for ($T_g$<1200K) from a rate coefficient determined experimentally at 298K [67]. According to this theory, the rate coefficient decreases with the increase of the gas temperature and is valid for gas temperatures below 1000K. We can verify that the transition probabilities P obtained are lower than 1 by calculating the reaction rates with the gas–kinetic collision frequency, obtained for a hard sphere model as [41],

$$\nu_{collision} = \sqrt{\frac{8 k_B T_g}{\pi \mu}} \pi R^2, \qquad (7)$$

with μ is the reduced mass of the colliding particles (taken as $3.65 \cdot 10^{-26}$ kg), R is the distance of collision assumed to be the Lennard Jones potential distance (taken as $3.763 \cdot 10^{-10}$ m [68]), $k_B$ is the Boltzmann constant and $T_g$ is the gas temperature. To avoid unphysical rate coefficients [17,20] (P>1), we use the same rate coefficient for all transitions with $v_3$>1.

*$CO_2$-O V-T*

The quenching of vibrationally excited $CO_2$ by O atoms is taken into account following the atmospheric model from Puertas *et al.* [69 70]. Two different mechanisms are considered here:

$VT_{3-O}$ : $CO_2(v_1 v_2^{l2} v_3 f)$ + O ↔ $CO_2(v_1(v_2=2,3,4)^{l2} v_3 f)$ + O with a rate coefficient, for $v_3$ = 1, given as:

$$k(cm^3 s^{-1}) = 2 \cdot 10^{-13} \cdot \left(\frac{T}{300}\right)^{1/2} \qquad (8)$$

$VT_{2-O}$ : $CO_2(v_1 v_2^{l2} v_3 f)$ + O ↔ $CO_2(v_1 v_2^{l2} -1 v_3 f)$ + O with a rate coefficient given as:

$$k(cm^3 s^{-1}) = \left(2.32 \cdot 10^{-9} \exp\left(-76.75 \cdot T^{-1/3} + 1 \cdot 10^{-14} \cdot T^{1/2}\right)\right) \qquad (9)$$

Both rate coefficients are scaled with a harmonic oscillator scaling (linear with v) for $v_3$ and $v_2 > 1$ according to [11].

*$CO_2(v)$-M and $CO(v)$-M V-T*

A possible way to obtain the $CO_2(v)$-M (M=CO, $O_2$) V-T rate coefficients is to multiply the known $CO_2(v)$-$CO_2$ coefficients by the constant 'relative efficiency' factor Φ suggested in [63]. However, a more general approach is to scale the rate coefficients according to the theoretical dependences from the SSH theory on the vibrational levels and gas temperature, as done in [41] for $N_2(v)$-$O_2$ and in [20] for $N_2(v)$-$CO_2$ energy transfers calculated from the V-T $N_2(v)$-$N_2$. The same approach was adopted for $CO(v)$-M (M=$CO_2$, $O_2$) V-T.

*CO-CO V-T*

The rate coefficients for the V-T transfers between CO molecules $CO(v)$ + CO ↔ $CO(v$-1) + CO taken from [71] were fitted for each vibrational quantum number with the gas temperature (K) using the following expression:

$$k(cm^3 s^{-1}) = T_g^p \cdot \exp\left(a T_g^{-1} + b T_g^{-1/3} + c T_g^{-2/3} + d T_g^{-4/3} + e T_g^{-5/3}\right) for\ 200 < T_g < 2000 \qquad (10)$$

with a, b, c, d, e and p fitting parameters depending on the vibrational level $v$ and $T_g$ the gas temperature and can be found in Table 1.

*Table 2: fitting parameters corresponding to the coefficients in expression () for the determination of the rate coefficient of the processes CO(v) + CO ↔ CO(v-1) + CO depending on the vibrational level v.*

| v | p | a | b | c | d | e |
|---|---|---|---|---|---|---|
| 1 | -15.7 | 1369066.1 | 7062.1 | -152282.5 | -5728820.0 | 9109363.3 |
| 2 | -17.4 | 1676528.2 | 8208.4 | -180552.3 | -7307224.3 | 12219240.7 |
| 3 | -17.4 | 1774248.6 | 8428.4 | -188018.3 | -7868692.9 | 13406893.1 |
| 4 | -16.8 | 1782710.7 | 8268.5 | -186764.9 | -7993304.5 | 13765067.3 |
| 5 | -16.0 | 1756422.0 | 7973.8 | -182242.0 | -7943995.6 | 13789509.0 |
| 6 | -15.1 | 1711488.3 | 7615.6 | -176048.2 | -7798579.8 | 13626437.5 |
| 7 | -14.1 | 1655326.2 | 7225.5 | -168912.3 | -7593034.4 | 13343357.2 |
| 8 | -13.2 | 1592228.9 | 6821.4 | -161251.1 | -7348313.0 | 12979633.9 |
| 9 | -12.3 | 1524967.7 | 6414.5 | -153330.1 | -7078047.8 | 12560968.5 |
| 10 | -11.5 | 1455436.6 | 6012.2 | -145327.9 | -6791643.7 | 12105186.1 |

The rate coefficients obtained through the fitting of the data from Cacciatore and Billing [71], as explained above, are in good agreement with the original data, as can be seen in Figure 1, especially between 300 K and 1000 K which corresponds to the range of gas temperatures measured for our conditions.

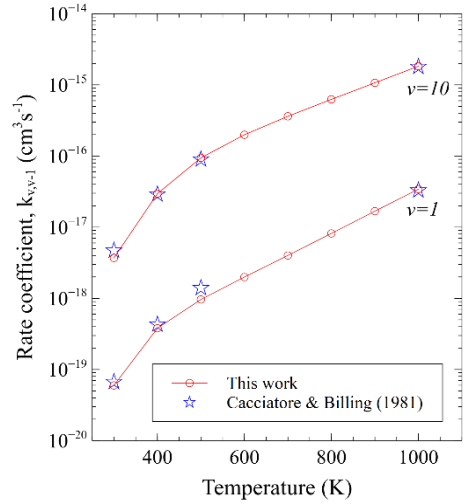

*Figure 1: Rate coefficients for the reaction CO(v) + CO → CO(v − 1) + CO as a function of the gas temperature. Data from [72] (☆) and fit (o).*

*CO-O V-T:*

The quenching of vibrationally excited CO by O atoms is included as described in [55], with an Arrhenius type temperature dependence:

$$k_{10}(CO-O)(cm^3 s^{-1}) = 5,3 \cdot 10^{-13} \cdot T_g^{1/2} \cdot \exp\left(-\frac{1600}{T_g(K)}\right) \quad (11)$$

and the harmonic oscillator scaling is assumed for $v > 1$.

*CO-CO V-V*

We include 90 V-V processes in our model for $0<v<10$ and we chose the results of the trajectory calculations from Cacciatore and Billing [72] as a reference. Indeed, the rate coefficients obtained from their calculations were in good agreement with experimental data between 100 and 500 K and for vibrational levels up to 10 or more. However, only 24 processes are present in [72] and we thus need to scale the missing rate coefficients. To do so, we used the FHO parametrization by Plönjes *et al.* [73] of the results obtained by Cacciatore and Billing [72]. Note that it is necessary to multiply the obtained rate coefficient by a factor $Z(cm^3 s^{-1}) = 3 \cdot 10^{-10} \left(\frac{T}{300}\right)^{1/2}$ to derive the overall rate coefficient.

To obtain a good agreement with the calculated rate coefficients from Cacciatore and Billing [72], we use the approximation of the adiabaticity factor given in [41] and we consider the characteristic length L=1·10$^{-11}$m similarly to the N$_2$-N$_2$ and N$_2$-O$_2$ systems where Guerra *et al.* [41] used L=2·10$^{-11}$m and L=3·10$^{-11}$m, respectively. We use a constant rate coefficient 8.85·10$^{-13}$cm$^{-3}$s$^{-1}$ taken from [72] (average of rate coefficients from 100K to 1000K) for the first process CO($v$=0) + CO($v$=1) → CO($v$=1) + CO($v$=0).

Finally, using the rate coefficient of the process CO($v$=1) + CO($v$=1) → CO($v$=0) + CO($v$=2) from [72] we renormalize the rate coefficient calculated from the procedure in Plönjes *et al.* [73]. We used a double exponential to fit the ratio between the rate coefficients from [72] and [73] dependent on the gas temperature and following the expression:

$$a \cdot \exp(b \cdot T) + c \cdot \exp(d \cdot T) \quad (12)$$

with a=2.088, b=-0.0018, c=0.08784 and d=4.1e-5.

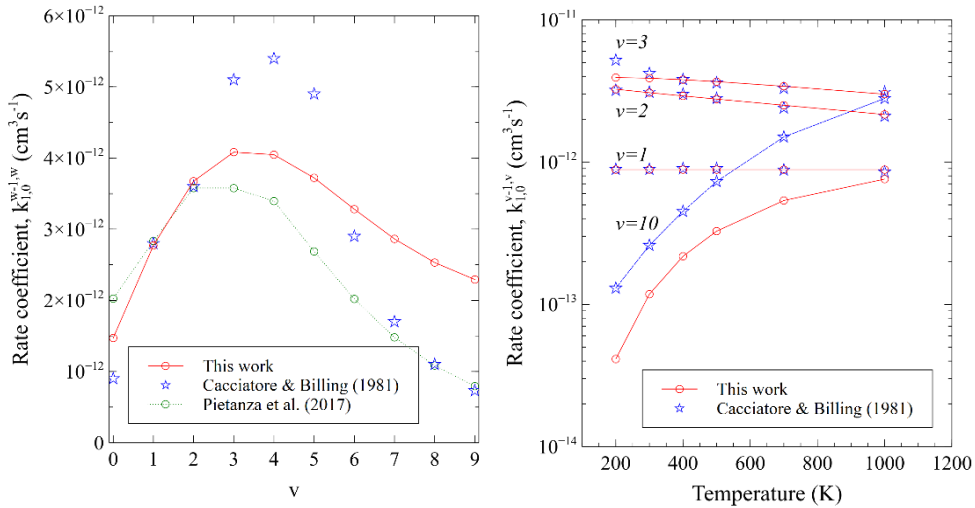

*Figure 2: On the left panel, rate coefficient for the non-resonant process: CO(1) + CO(v-1) → CO(0) + CO(v) at 500K as a function of v and one the right panel rate coefficient for the same process as a function of the gas temperature, compared with the results from Cacciatore and Billing [76] and Pietanza et al. [74].*

In Figure 2, the rate coefficients for the non-resonant process, $CO(1) + CO(v-1) \rightarrow CO(0) + CO(v)$, are plotted as a function of v, at 500K, and compared with the results from Cacciatore and Billing [72] and Pietanza *et al.* [74]. The variation of the other available rate coefficients with the gas temperature was also verified and it is satisfactory even for higher vibrational levels.

*$CO_2$-CO V-V*

As pointed out in the introduction, the transfers between vibrationally excited CO and the asymmetric stretch mode of $CO_2$ are very efficient and can promote the ladder climbing mechanism along this $CO_2$ mode, with a potential positive effect on $CO_2$ dissociation. CO molecules can transfer a considerable amount of energy to the $v_3$ vibration because the difference between the energies of the first vibrational level of CO and the $(00^011)$ level of $CO_2$ is only 170cm$^{-1}$, which is smaller than the average kinetic energy kT [75]. Moreover, the observed cross sections of excitation of molecular vibrations of CO are unusually large, which is related to the resonance effect of short-lived negative ions CO$^-$ [76].

Kustova *et al.* [77] provide a straightforward procedure to obtain accurate rate coefficients for the V-V transfer between $CO_2$ and CO.

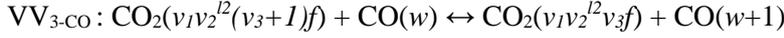

$VV_{3\text{-}CO}$ : $CO_2(v_1v_2{}^{l2}(v_3+1)f) + CO(w) \leftrightarrow CO_2(v_1v_2{}^{l2}v_3f) + CO(w+1)$

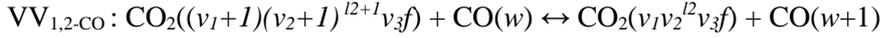

$VV_{1,2\text{-}CO}$ : $CO_2((v_1+1)(v_2+1)^{l2+1}v_3f) + CO(w) \leftrightarrow CO_2(v_1v_2{}^{l2}v_3f) + CO(w+1)$

The rate coefficients of vibrational energy transitions between the lowest vibrational states are computed using experimental data [78] and can be calculated using the expression:

$$k_{0\rightarrow 1} = \frac{kT}{P\tau}, \text{with } P\tau = 10^{A_0 + A_1 T^{-1/3} + A_2 T^{-2/3}} \tag{13}$$

The value of the $A_n$ constants can be found in table 1 of [77]. The remaining rate coefficients (for higher states) are calculated on the basis of the harmonic oscillator modified for polyatomic molecules.

For $VV_{3\text{-}CO}$, we use the scaling $k(w \rightarrow w+1)(v_3 + 1 \rightarrow v_3) = k_{0\rightarrow 1} * (v_3 + 1) * (w+1)$

For $VV_{1,2\text{-}CO}$, we use the scaling $k(w \rightarrow w+1)(v_{1,2} + 1 \rightarrow v_{1,2}) = k_{0\rightarrow 1} * (v_1 + 1) * (v_2 + 1) * (w+1)$.

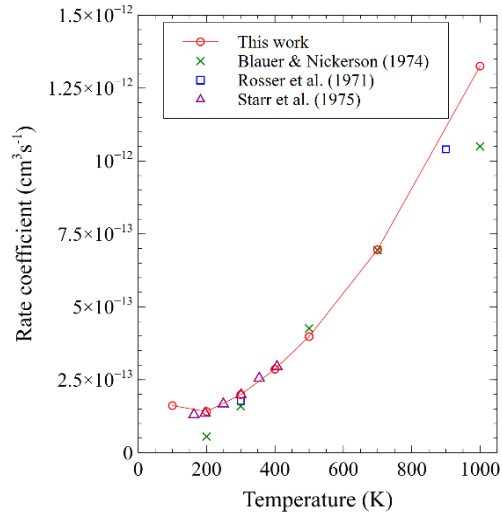

*Figure 3: rate coefficients for the process: $CO_2(00^011) + CO(0) \rightarrow CO_2(00^001) + CO(1)$ used in this work (o) against the results from Rosser et al. [79], Starr et al. [80] and Blauer and Nickerson [63] (symbols).*

We compared the values obtained in this work following the procedure from Kustova *et al.* [77] for the process $CO_2(00^011) + CO(0) \rightarrow CO_2(00^001) + CO(1)$ with values determined experimentally by Rosser [79] *et al.* (linear dependence between 300 and 900K within experimental error), Starr *et al.* [80] and Blauer and Nickerson [63] and found a good agreement, as shown in Fig. 3.

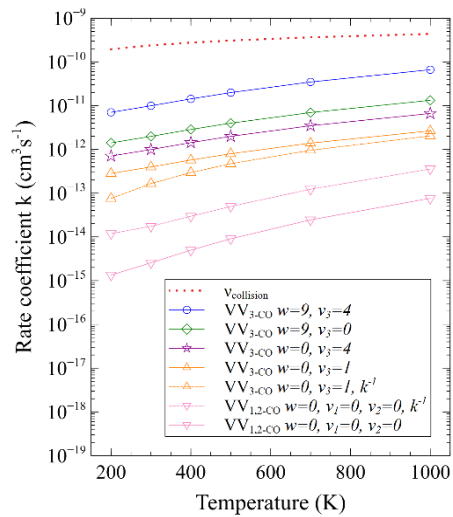

*Figure 4: rate coefficients for the direct and indirect ($k^{-1}$) processes : $CO_2(v_3+1) + CO(w) \leftrightarrow CO_2(v_3-1) + CO(w+1)$ ($VV_{3-CO}$) and $CO_2(v_1+1,v_2+2) + CO(w) \leftrightarrow CO_2(v_1-1,v_2-1) + CO(w+1)$ ($VV_{2-CO}$), and gas collision frequency $\nu_{collision}$ as a function of the gas temperature for different combination of $v_1$, $v_2$, $v_3$ and w (being equal to zero if not specified in the legend).*

The rate coefficients of the reverse transitions are related to the rate coefficients of forward transitions by the detailed balance principle. The collision frequency (or gas kinetic rate) represented in Figure 4 along with a few calculated rate coefficients, is estimated via a hard sphere model (c.f. expression 7), gives an upper limit for the scaled rate coefficients and we ensure that no rate coefficient corresponds to a probability above 1.

*Summary of the most important rate coefficients.*
The rate coefficients of vibrational energy transitions between the lowest states of several V-V and V-T processes are plotted in Figure 5. The rate coefficients of these processes differ by several orders of magnitude. The most important ones in term of amplitude are the quenching of vibrationally excited $CO_2$ and CO molecules by atomic oxygen ($VT_{3-O}$ and $VT_{CO-O}$) and the vibrational-to-vibrational transfer between two $CO_2$ molecules ($VV(SB)_{CO2-CO2}$), two CO molecules ($VV_{CO-CO}$) and finally the vibrational transfer between CO and $CO_2$ ($VV_{3-CO}$).

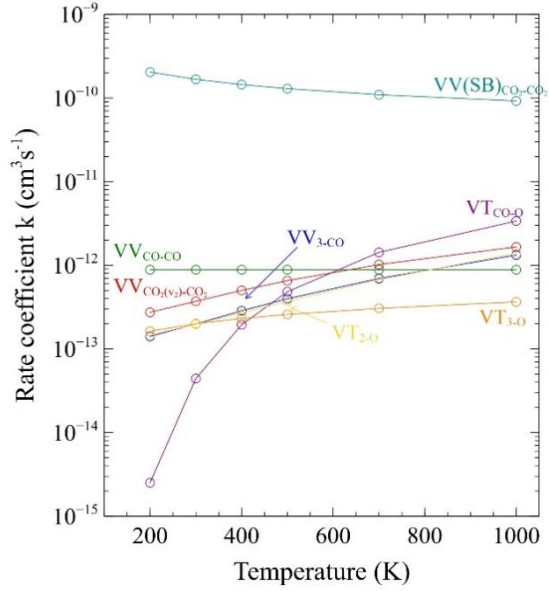

*Figure 5: Probabilities of vibrational energy transitions between the lowest levels.*

## IV. Results and discussion

The results of the model are compared with experimental measurements of the reduced density of atomic oxygen, reduced field, dissociation fraction and vibrational temperatures of $CO_2$ and CO in a DC reactor at pressures between 0.4 and 5 Torr (53.33 - 666.66 Pa). This comparison provides the validation of the model, as well as the interpretation of the measured quantities and the identification of the main processes ruling the discharge.

It is worth noting that our state-to-state model provides the populations of each individual vibrational levels of the different modes of $CO_2$ and of CO. Therefore, the vibrational temperature is calculated, assuming a Treanor distribution [81], as:

$$T_{v,ij} = \left( \frac{E_1}{\ln(p_i/p_j) - \dfrac{E_j - E_i - E_1}{k_B T_g}} \right) / k_B \qquad (15)$$

where $E_1$ is the energy of the first level, $p_i$ and $p_j$ and $E_i$ and $E_j$ are the population and energy of level i and j, respectively, $T_g$ is the gas temperature and $k_B$ the Boltzmann constant. We use an average of the temperatures calculated using the first three vibrational levels.

Moreover, since the characteristic temperatures corresponding to the effective symmetric mode, $T_1$, (which includes the Fermi resonant states) and bending mode, $T_2$, in our conditions, are nearly the same, we define a common temperature of the bending and symmetric stretching modes denoted $T_{1,2}$ [18,34]. This comes from the occurrence of the Fermi resonance between the symmetric and bending modes of vibration and with the similarity of the energies and rate coefficients involving Fermi and non-Fermi bending levels. Therefore, the two characteristic temperatures $T_{1,2}$ and $T_3$ are enough for a simple description of the extent of vibrational excitation although this is not imposed in the model.

   1) Validation of the CO vibrational kinetics and effect of the wall deactivation

This section focuses on the validation of the CO vibrational kinetics by comparing the simulations results with experimental data obtained in a DC glow discharge ignited in 100% $CO_2$ and a current of 40mA [34].

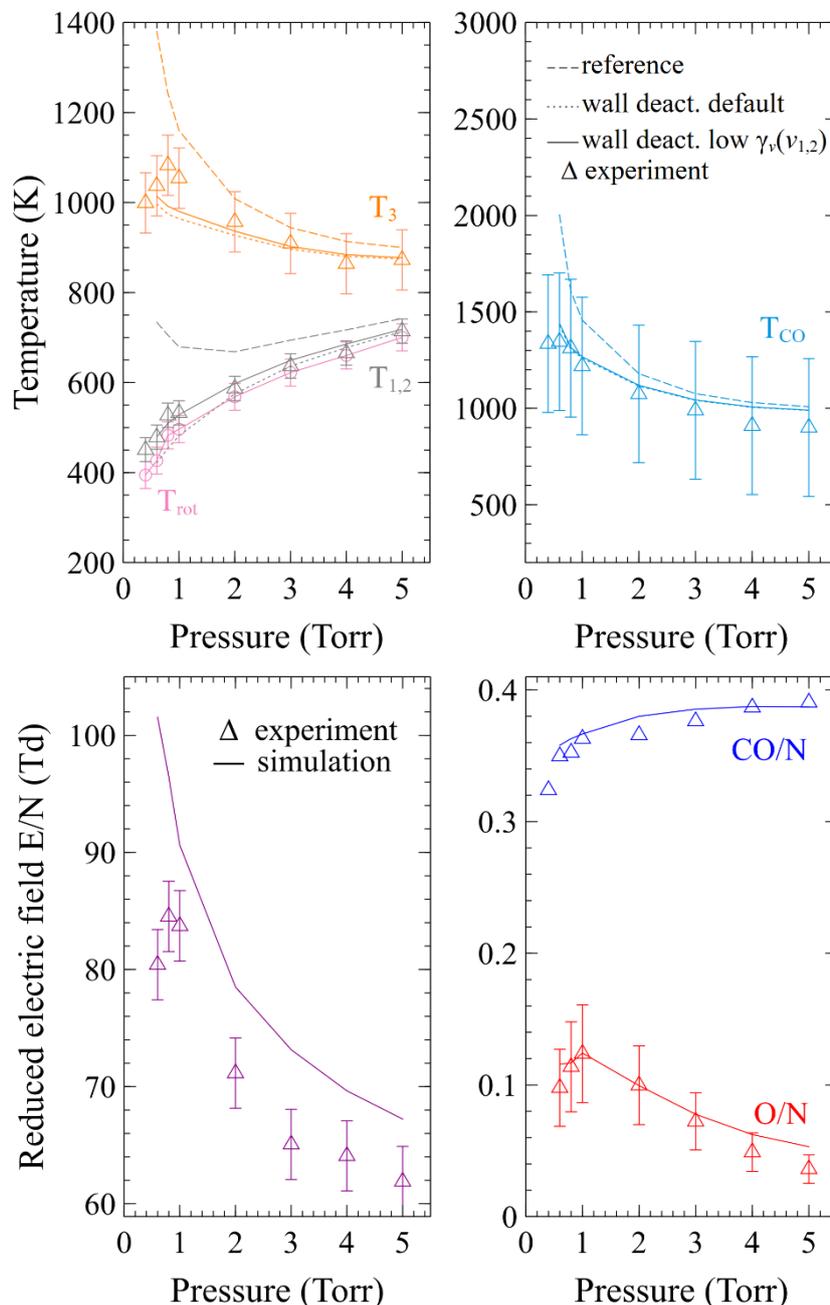

*Figure 6: Experimental values (Δ) and calculated values (line) of the common vibrational temperature of the $CO_2$ bending and symmetric modes $T_{1,2}$, the vibrational temperature of the asymmetric stretching mode $T_3$, the CO vibrational temperature $T_{CO}$, the rotational temperature $T_{rot}$, the reduced electric field E/N, the CO reduced density CO/N and the reduced atomic oxygen density O/N, when a discharge is ignited in $CO_2$, at current = 40 mA and as a function of pressure. The model calculations were done including with the default probabilities from section III.3. (···), $\gamma_v(v_{1,2}) = 0,05$ (—) and excluding (– –) the wall deactivation of the vibrationally excited states of CO and the different modes of $CO_2$, at the wall.*

In Figure 6 the measured and calculated common vibrational temperature of the $CO_2$ bending and symmetric modes $T_{1,2}$, the vibrational temperature of the asymmetric stretching mode $T_3$, the CO vibrational temperature $T_{CO}$, the rotational temperature $T_{rot}$, the reduced electric field and the reduced densities of atomic oxygen and CO are presented. These two last quantities show a very good agreement between calculations and measurements and the self-consistently calculated reduced electric field is overestimated for all conditions and a few reasons for this discrepancy are discussed in section IV.3. The CO vibrational kinetics scheme added to the model is able to successfully reproduce the measured $T_{CO}$ values within an error of 10%, we can therefore consider that our CO vibrational kinetic scheme is validated. Overall, all calculated quantities are in very good agreement with the experimental data and more specifically, for the temperatures, when the wall deactivation of vibrations is included.

The influence of deactivation of vibrationally excited $CO_2$ and CO at the walls, described in section 3, is illustrated in Figure 6, for the case of pure $CO_2$. Including the wall deactivation for vibrationally excited $CO_2$ and CO molecules mostly affects the vibrational temperatures of these molecules but not the other quantities like O/N, CO/N and E/N. $T_{CO}$, $T_{1,2}$ and $T_3$ decrease for all conditions and the trends as a function of the pressure are improved and leading to a better agreement with the measurements. The results change drastically in the case of a discharge below 2 Torr, while they are very similar at higher pressures, as expected, and also observed in [20]. Moreover, we could observe that including the detailed balance for the wall deactivation processes plays an important role, up to 7 % increase of $T_{1,2}$, for the lowest pressure. The default deactivation probabilities of the different modes of $CO_2$ lead to calculated $T_{1,2}$ values lower than in the experiment, we thus propose to reduce the deactivation probability for the bending, symmetric stretch, and mixed modes from 0.2 to 0.05 and this corresponds to the results labelled $\gamma_v(\nu_{1,2}) = 0{,}05$ in Figure 6.

2) Effect of atomic oxygen on the vibrational kinetics of $CO_2$ and CO

In order to study the effect of the quenching of vibrational energy by O atoms as a function of the $O_2$ content we have reproduced with our model the experimental conditions discussed in [33]. These conditions are essentially the same as described in this work. However, the inside of the Pyrex tube was covered with a layer of micro-structured silica fibers, increasing the effective surface area in contact with the plasma to enhance the O atom recombination and consequently reduce the density of O atoms in the gas mixture. The vibrational and rotational temperatures of $CO_2$ and CO were measured with in situ FTIR spectroscopy and atomic oxygen density and loss frequency by actinometry. In our simulations we can reproduce this experiment by setting the recombination probability $\gamma_O$ to 1 and the tube length to 23 cm. Note that as a first estimation we assume a constant value of gamma ($\gamma_O=1$) to capture the phenomenon, however a dependence with pressure could improve the agreement between simulation and experiment.

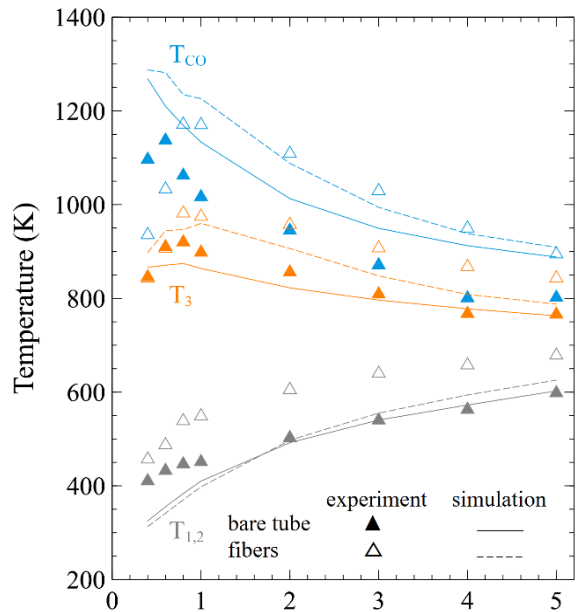

*Figure 7: Experimental values (Δ) and calculated values (line) of the common vibrational temperature of the $CO_2$ bending and symmetric modes $T_{1,2}$, the vibrational temperature of the asymmetric stretching mode $T_3$ and the CO vibrational temperature $T_{CO}$, for a pure $CO_2$ discharge, at current= 20 mA with O loss probability $\gamma_O$ obtained experimentally [32] with a bare Pyrex tube (—) and $\gamma_O=1$ (– –) used in the calculations.*

Figure 7 presents the vibrational temperatures calculated in the simulations and measured experimentally, in the case of the bare tube, corresponding to $\gamma_O$ obtained from loss frequency measurements in [34] and for the tube covered with fibers, corresponding to $\gamma_O=1$ in the simulations. As can be observed in Fig. 7, the different vibrational temperatures, from the calculations and experiments, as a function of the pressure, are in good agreement. Moreover, the fact that the calculated $T_{1,2}$ is too low when compared with the experiments could come, at least partly, from the wall deactivation for which we use the default value presented in section III.3. As discussed above using a lower value of wall deactivation probability and including the reverse wall deactivation processes according to the detailed balance led to better results. Besides, the same wall deactivation was considered for both conditions (with and without fibers); however, changing the surface conditions of the plasma reactor can also increase the deactivation of vibrationally excited molecules at the surface.

The experimental results [33] show a remarkable increase of the vibrational excitation of both $CO_2$ and CO with the large surface material confirming that atomic oxygen is a strong quencher of the vibrations of both species. Besides, it was also verified that neither the dissociation fraction nor the reduced electric field were changing significantly, within the reproducibility error and that O atom density decreased drastically, down to ~5% of the density measured with the bare tube.

Likewise, in the simulations, the dissociation fraction is only changing by around 0.05 and the reduced field by less than 2 % while the reduced atomic density goes down to 0.7 % - 2.5 % (depending on the pressure) of the reference O/N attesting than atomic oxygen does not participate in back reactions as stated in [33]. Besides, the increase of $T_3$ by 100K (Fig. 7), in average, does not influence much the dissociation fraction and this proves that the vibrational kinetics are not playing a significant role in the dissociation of $CO_2$. Therefore, the quenching of the $CO_2$ vibrations by O atoms, which becomes more and more important as the $O_2$ fraction increases in the mixture, cannot explain the detrimental effect of $O_2$ addition on $CO_2$ dissociation seen in Figure 10. However, this study illustrates the role of O atoms in the quenching of vibrations which is essential in these conditions and should not be overlooked.

### 3) Addition of $O_2$

The admixture of $O_2$ has a detrimental impact on $CO_2$ decomposition since it leads to a decrease of the dissociation fraction defined as:

$$\alpha = \frac{n_{CO}}{n_{CO} + n_{CO_2}} \qquad (16)$$

This effect was already observed experimentally by Grofulović *et al.* [12] and confirmed in the present study (Fig. 10). Two main reasons were discussed briefly in the introduction, one of them being the enhancement of the reverse reaction producing back $CO_2$ in the presence of $O_2$ [13]. Another possible explanation is the quenching of vibrationally excited $CO_2$ by O atoms. However, according to the previous section the decrease of the $CO_2$ dissociation as a function of the $O_2$ content is related to the back reactions with $O_2$ and not to the quenching of vibrational energy by O atoms. The study of $CO_2-O_2$ gas mixtures gives further insight into the impact of the oxygen content in these two mechanisms.

Figures 8 and 9 show, respectively, the measured and calculated values of the reduced electric field, E/N, and the reduced atomic oxygen density, O/N, as a function of the $CO_2$ initial fraction for a discharge current of 20 mA and 40 mA and for two different pressures with N the density calculated from the ideal gas law with the pressure and gas temperature taken from the FTIR measurements.

Note that the data for 100% $CO_2$ is consistent with previously measured data in similar conditions [34] (see previous section).

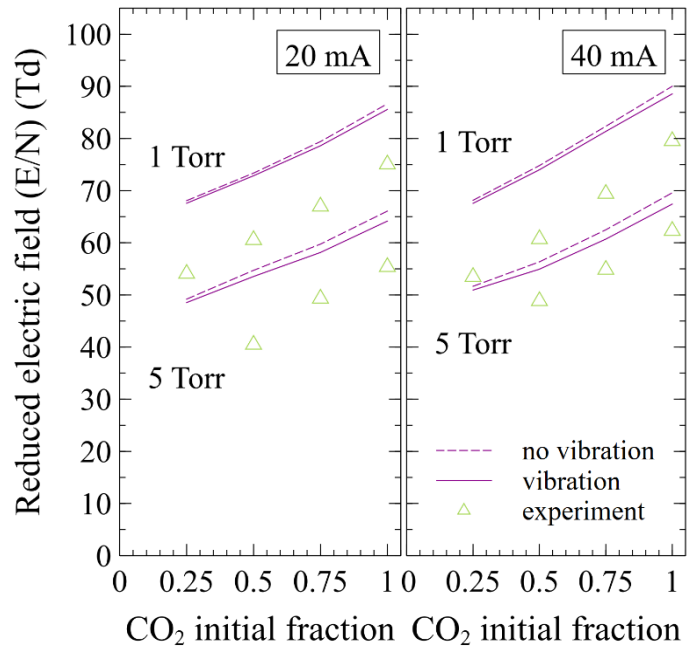

Figure 8: Reduced electric field E/N, of a $CO_2$-$O_2$ discharge as a function of the $CO_2$ initial fraction, at a current of 20 mA and 40 mA for 1 and 5 Torr: experiment (Δ), model calculations excluding (– –) and including (—) the vibrational kinetics in Loki C.

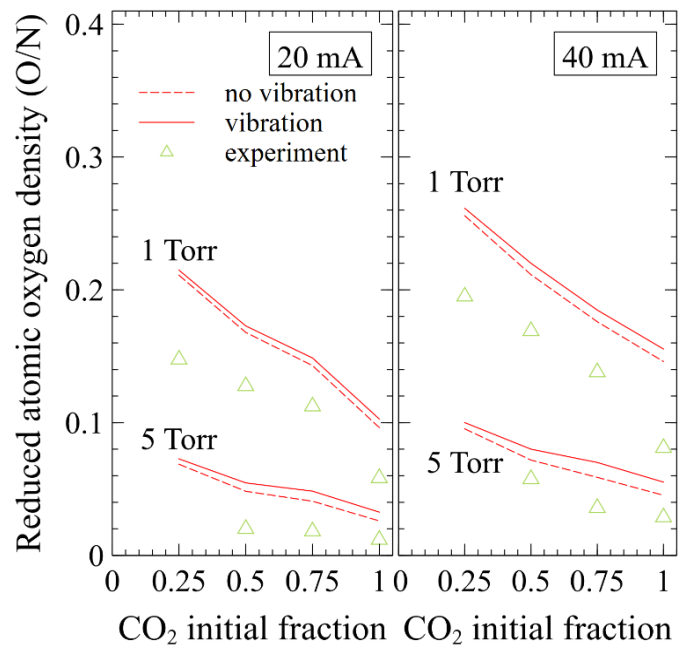

Figure 9: Reduced atomic oxygen density O/N of a $CO_2$-$O_2$ discharge as a function of the $CO_2$ initial fraction, at a current of 20 mA and 40 mA for 1 and 5 Torr: experiment (Δ), model calculations excluding (– –) and including (—) the vibrational kinetics in Loki C.

Overall, the variation of the self-consistently calculated reduced electric field with the $CO_2$ initial fraction for different pressures agrees fairly well with the experimentally measured E/N as shown in Figure 8. These results prove that the ionization rates and ion chemistry (transport and charge exchange) are well characterized and that the model can be used as a predictive tool when no experimental data for E/N are available. However, discrepancies are still present and further investigation is required to clarify why the absolute value of E/N seems overestimated for all conditions. Different possibilities for improvements of the model were already given in [11] and concern the rate coefficients of several reactions involving charged species, stepwise ionization processes involving vibrationally and electronically excited CO and $CO_2$ molecules, the charged-particle transport model and the ion transport data. Using the effective diffusion scheme for charged-species transport rather than classical ambipolar diffusion gave a better agreement between calculated and experimental E/N in [14]. However, this study was done in pure $CO_2$ were the low electronegativity observed did not invalidate the use of effective ambipolar diffusion which is not the case anymore in the $CO_2$-$O_2$ mixture.

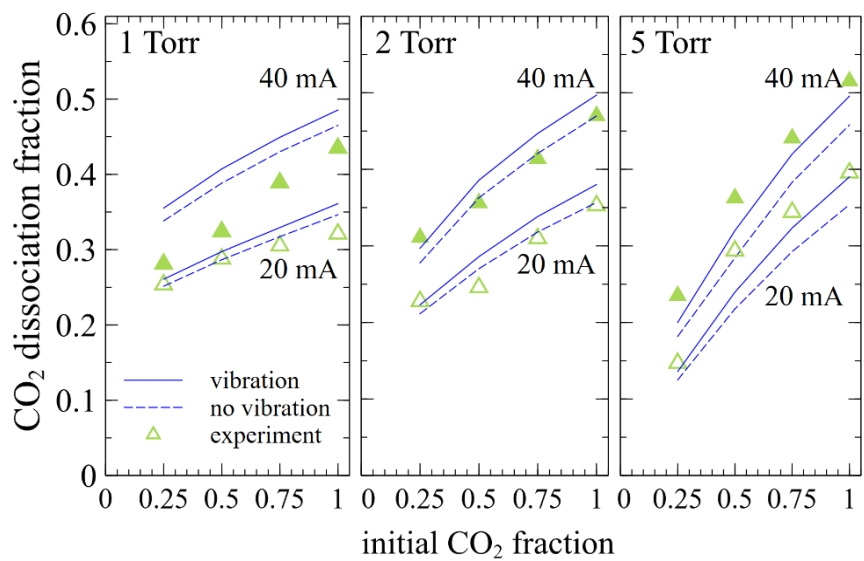

*Figure 10: $CO_2$ dissociation fraction of a $CO_2$-$O_2$ discharge as a function of the $CO_2$ initial fraction in the $CO_2$-$O_2$ mixture at 20 mA and 40 mA: experiment (Δ), model calculations by excluding (– –) and including (—) the vibrational kinetics. No error bars are included in this figure, since the fitting error of the FTIR spectra and the reproducibility error of the experimental results are smaller than the size of the symbols.*

The reduced atomic oxygen density, O/N, (Fig. 9) also shows a good trend with the $O_2$ fraction but remains too high for all conditions. This discrepancy could be reduced by using Polak's cross sections for the $O_2$ dissociation by electron impact [25] and it is discussed in section IV.5.

In Figure 10 we present the $CO_2$ dissociation fractions as a function of the gas mixture at different pressures and currents when the vibrational kinetics of $CO_2$ and CO are excluded or included in the model. For all the conditions studied, the dissociation fraction $\alpha$ increases when the vibrational kinetics are considered. This can be a result of modifications of the EEDF or because of the contribution of the vibrationally excited states of $CO_2$ to the dissociation by electron impact. Indeed, the rate coefficients for the latter process are higher from the vibrationally excited states than from the ground state, due to the threshold shift in the dissociation cross sections for the vibrationally excited levels of $CO_2$. Moreover, the high energy tail of the EEDF can be enhanced due to superelastic collisions with vibrationally excited CO and $CO_2$. The effect of the vibrational populations of CO and $CO_2$ in the electron kinetics on the EEDF was already thoroughly studied for pure $CO_2$ plasmas in [11,14]. The vibrational populations determine the rate of superelastic collisions with the vibrational levels and can therefore modify the high energy-tail of the EEDF increasing, by orders of magnitude, the electron impact excitation (including dissociation) and ionization rates [82-84]. In the system under study, $CO_2$ is essentially dissociated by direct electron impact, both on molecules in the vibrational ground-state $(00^001)$ and in vibrationally excited states. The contribution of the latter states comes mainly from the lower-laying levels $(01^101)$, $(02^201)$ and $(10^002)$. At 1 Torr, and in pure $CO_2$, which corresponds to the condition of highest $T_{1,2}$ and $T_3$, 85% of the dissociation occurs from the ground state (GS). Moreover, for this case, the population of $CO_2$ in the GS is 0.67 and the corresponding dissociation rate coefficient from the $(01^10)$ and the $(02^20)$ levels is less than twice the rate coefficient from the GS. We thus conclude that, in our conditions, superelastic collisions have a prominent role on the enhancement of the $CO_2$ dissociation.

To investigate the effect of the discharge current on different plasma parameters, we performed simulations at 20 mA and 40 mA. The conversion is strongly correlated to the discharge current and the dissociation fraction, α, is increased by ~0.11, both in the simulations and experiment when increasing the current from 20mA to 40mA (Fig. 10). Indeed, increasing the current changes the electron density in an almost linear way and in turn the electrons participate to the dissociation of $CO_2$. It was also verified that the experimental and calculated $T_{1,2}$, $T_3$ and $T_{CO}$ increase by a similar amount both in the experiment and simulations when the current is doubled.

At low pressure the experimental values are closer to the case without vibrations and always lower than the model. At 2 Torr, we can see that the agreement improves and finally at the highest pressure the experimental values are closer to the calculations performed with vibrations. This suggests that at higher pressure it is more important to take into account the vibrational kinetics to describe the chemistry in the discharge and at lower pressure the effect of the walls (both directly in the chemistry in surface reactions and in the vibrational kinetics) is not fully reproduced by the model.

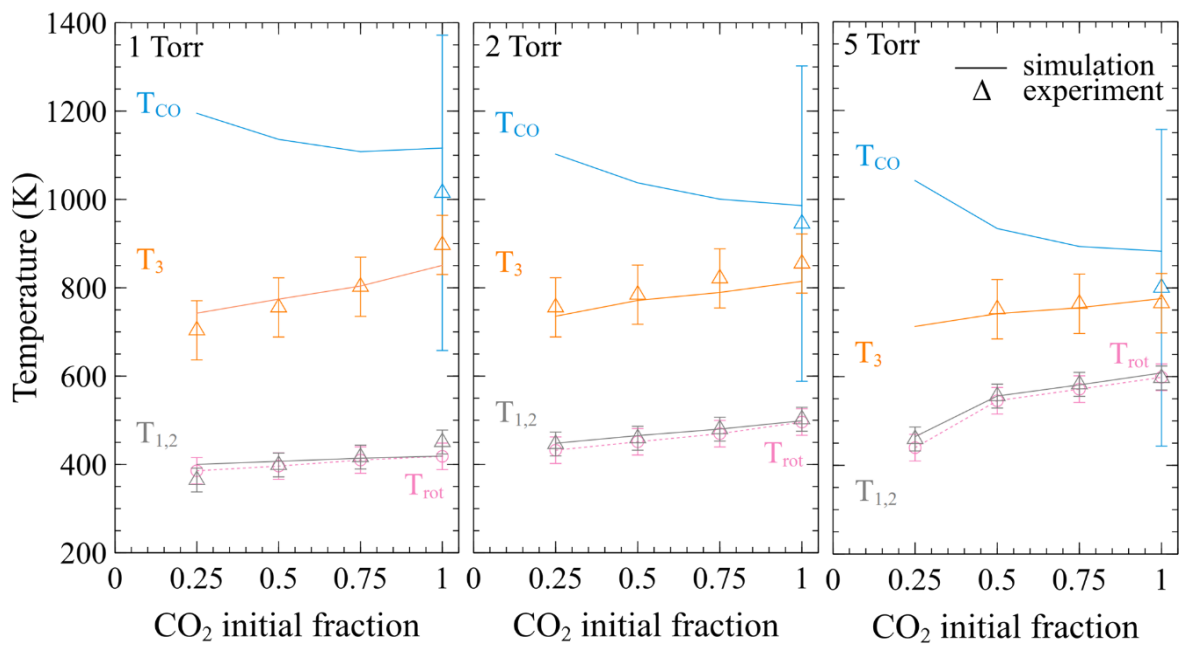

*Figure 11: Experimental values (Δ) and calculated values (line) of the common vibrational temperature of the $CO_2$ bending and symmetric modes $T_{1,2}$, the vibrational temperature of the asymmetric stretching mode $T_3$, the CO vibrational temperature $T_{CO}$ and the rotational temperature $T_{rot}$ (used as input parameter for the model) when a discharge is ignited in different mixtures of $CO_2$-$O_2$, at current= 20 mA, and pressures of 1, 2 and 5 Torr. The error bars indicated were obtained in pure $CO_2$ at 5 Torr and 50mA [26].*

Figure 11 shows the vibrational temperatures for CO and for the different modes of $CO_2$ obtained from the model and experiment as well as the rotational temperature $T_{rot}$ used as input parameter for the model. They are in very good agreement and the discrepancy for $T_{CO}$, especially at 1 Torr comes from the signal to noise ratio worsening because of the lower density in general and because the dissociation is relatively low. For all the conditions under study, $T_{1,2}$ is almost in equilibrium with $T_{rot}$ ($\sim T_g$) but $T_{CO}$ is higher than $T_3$ which is higher than $T_{1,2}$. $T_{CO}$ showing larger values than $T_3$ can be explained by different rate coefficients for V-T relaxation, the lack of inter-mode V-V relaxation processes affecting $CO_2(v_3)$ ($CO_2(v_3)+CO_2(v_{1,2}) \rightarrow CO_2(v_3-1)+CO_2(v_{1,2}+1)$) but not CO, and more efficient vibrational excitation through electron-to-vibrational energy transfers for CO [34].

We can see in Figure 11 that the temperatures $T_3$ and $T_{1,2}$ are slightly decreasing with increasing $O_2$ content. Indeed, the atomic oxygen density is higher when the $O_2$ content increases (Fig. 7) and therefore the quenching of the $CO_2$ vibrations by O atoms, which is an efficient process, is more important. We can also notice an opposite trend for the $T_{CO}$, that can be explained by the dependency of the rate coefficient for the $CO_2$-CO V-V transfer on $T_g$. Indeed, $T_g$ decreases with the $O_2$ proportion which leads to a smaller rate coefficient for this process. Moreover, the dilution of the $CO_2$ and CO molecules when $O_2$ is added can also explain the trend observed. Indeed, on the one hand the collisions and therefore vibrational energy transfer between CO and $CO_2$ are reduced because these molecules are diluted. On the other hand, $CO_2$ molecules will have more collisions with $O_2$ molecules with whom the vibrational energy transfer is much less efficient than with CO molecules. These two effects contribute to reduce $T_3$ and $T_{1,2}$ and increase $T_{CO}$ because of the reduced CO-$CO_2$ V-V (in addition to the effect of $T_g$). Note that the rate coefficient associated with the de-excitation of CO by $O_2$ is also increasing as a function of $T_g$ but this increase is very slow in comparison with that of $CO_2$ and the absolute value is two orders of magnitude lower than for $CO_2$.

Finally, regarding the effect of pressure, we observe a smaller difference between the temperatures $T_3$ and $T_{CO}$ (and $T_3$ and $T_{1,2}$) as pressure increases. This comes from the vibrational energy transfer from CO to $v_3$ of $CO_2$, occurring due to its near resonant frequency, more effective at higher pressures because of the higher collision frequencies.

4) Effect of the first electronically excited state of CO on the $CO_2$ dissociation and recombination

The role of the electronically excited state $CO(a^3\Pi_r)$, hereafter denoted CO(a), on $CO_2$ dissociation can be beneficial or detrimental for the $CO_2$ conversion. In fact, CO(a) can have an ambivalent role depending on the CO and $O_2$ density [85] as it either enhances the dissociation of $CO_2$ or stimulates the reconversion back to $CO_2$. Cenian *et al.* [85] simulated glow discharges with similar working conditions to ours and brought up the ambivalent role of CO(a) and stressed its importance in the full description of $CO_2$ decomposition. Indeed, despite having a small molar fraction (~$10^{-7}$), similar to what was reported in [14, 85], the energy of this state (~6 eV) is enough to dissociate $CO_2$ and $O_2$ molecules and, owing to this high energy, the rate coefficients of the processes involving CO(a) are close to the gas kinetic collision frequencies.

While there are several possibilities for back-reaction mechanisms involving CO and $O_2$: reactions between ground-state molecules and reactions involving vibrationally or electronically excited CO, the experimental results and preliminary calculations by Morillo-Candas *et al.* [13] and the kinetic modelling by Silva *et al.* [31] show a key role of the metastable electronically excited state $CO(a^3\Pi_r)$ in the back-reactions, in low pressure pulsed glow and RF discharges. Indeed, the recombination of CO and $O_2$ both in the ground states producing $CO_2$ is a possible 'back reaction' but very slow at room temperature [86] and it is not even included in our model. This rate coefficient becomes significantly higher if the reaction involves vibrationally [26] or electronically excited CO molecules [13, 87]. Back-reactions based on the vibrationally excited CO were not dominant in their discharge conditions [13] but could become relevant at slightly higher vibrational temperatures, like in microwave discharges. Since the vibrational temperatures in those works are similar to the ones in our discharge, we can assume that vibrationally excited CO does not play an important role, in our conditions, for the back reactions.

In order to assess the role of CO(a) in the present conditions, Figure 12 compares the simulations when this state is included or excluded from the model. We can observe that, when CO(a) is added in the model, the dissociation fraction increases for $CO_2$ initial fraction of 0.75 and above, while below this turning point the dissociation fraction decreases. For gas mixtures with large amount of $CO_2$ but low CO density, the reaction $CO(a) + CO_2 \rightarrow 2CO + O$ contributes to enhance the dissociation. On the contrary, if the concentrations of CO and $O_2$ are larger the processes $CO(a) + O_2 \rightarrow CO_2 + O$ and $CO(a) + CO \rightarrow CO_2 + C$ are prevailing and lead to the $CO_2$ reconversion [13, 31, 85]. The addition of $O_2$ can also modify the ion conversion pathways and induce changes in the plasma parameters like the gas temperature [12].

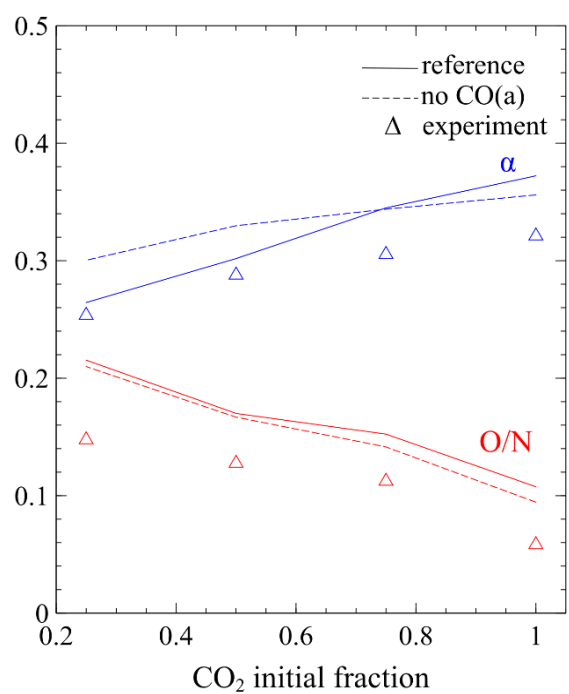

*Figure 12: Experimental values (Δ) and calculated values (line) of the dissociation fraction α and the reduced atomic oxygen density O/N, for a discharge ignited in different $CO_2$-$O_2$ mixtures, at 20 mA and 1 Torr, including (—) and excluding (– –) the CO(a) state from the simulations.*

The lack of experimental data for branching ratios of the different dissociative quenching mechanisms of the CO(a) state has an impact on the simulated dissociation fraction, as pointed out in [14]. Furthermore, some energy transfer processes between vibrational CO($v$) and CO(a), not included in the model, can take place in the system. For instance, the quenching of CO(a) via $CO(a) + CO \rightarrow 2CO(v)$ [74, 88] can produce vibrationally excited CO, while collisions between sufficiently energetic vibrational states can lead to CO(a) formation in $CO(v) + CO(w) \rightarrow CO(a) + CO$ [89]. Another possibly important process is the pumping of energy in the $v$=27 level by quenching of CO(a), $CO(a) + CO \rightarrow CO(v=27) + CO$ [74]. Finally, one can also consider the formation of the metastable through the following reaction $CO(v>27) + CO \rightarrow CO(a) + CO$ [90]. These mechanisms, although possibly important for the chemistry, were not included in this work as they are not likely to affect the results and conclusions in the steady-state conditions under study and would require an extended description of the CO vibrational kinetics. An assessment of the relevance of these energy transfer processes will be investigated in the future.

5) $O_2$ dissociation cross sections

Two reactions account for $O_2$ dissociation by electron impact:

$$e + O_2 \rightarrow e + O + O \ (6\ eV) \tag{15}$$

$$e + O_2 \rightarrow e + O + O(^1D) \ (8.4\ eV) \tag{16}$$

The dissociation through channel (15) occurs via the Herzberg states $O_2(A^3\Sigma_u^+, C^3\Delta_u, c^1\Sigma_u^-)$ and gives two oxygen atoms in the ground state. The oxygen dissociation corresponding to channel (16) occurs via the excitation of the $O_2(B^3\Sigma_u^+)$ state continuum and one of the oxygen atoms produced is in an electronically excited state $O(^1D)$. The continuum excitation of the $O_2(B^3\Sigma_u^+)$ state is usually the main contributor to the total cross-section of oxygen dissociation through electron impact. However, near the dissociation threshold the main contribution is made by the excitation of the Herzberg states with the energy threshold of around 6 eV. As indicated in the model description our default cross sections for $O_2$ electron impact dissociation are taken from [44, 57]. However, it was reported that these cross sections may be overestimated and that it may be necessary to reduce the contribution from process (15) [91]. Thus, Kovalev *et al.* used modified electron impact cross-sections for oxygen dissociation channels (15) and (16) as presented in [92] with respective thresholds of 5.58 and 7.34 eV. This modified cross-section set was verified by comparison with a large set of experimental data in different oxygen discharges [93, 94]. Other dissociation cross sections with lower amplitudes can be found in the literature. For instance, Polak and Slovetsky [25] computed the electron impact dissociation for cross-sections of $O_2$ and verified that the calculated cross-section of dissociation from the levels of the $O_2(B^3\Sigma_u^+)$ state was in satisfactory agreement with a few experimental points. Laporta *et al.* [95] calculated a cross section for resonant electron impact dissociation of oxygen and Itikawa [96] reported a cross section for the total dissociation of $O_2$ in neutral products. The cross sections mentioned above are represented in Figure 13.

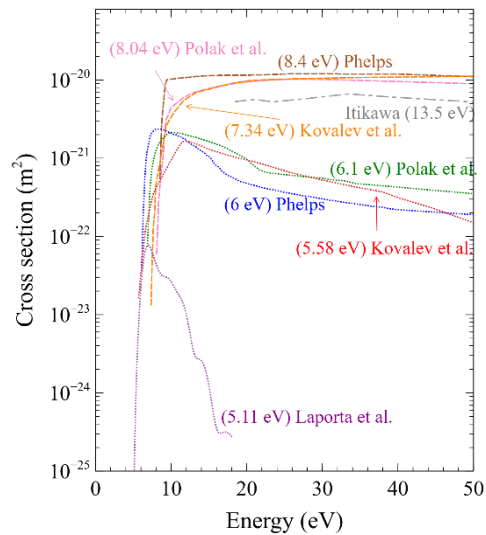

*Figure 13: Dissociation cross sections of $O_2$, $e + O_2 \rightarrow O + O$ (···), $e + O_2 \rightarrow O + O(^1D)$ (– –), total dissociation (– - –) from different references [25, 57, 92, 95, 96] and the corresponding thresholds in parenthesis.*

In order to assess the influence of the $O_2$ electron impact dissociation cross section we make additional calculations by replacing the cross sections from Phelps [57] by the ones from Polak and Slovetsky [25]. However, similarly to what was done for the electron-impact dissociation of $CO_2$, we only use the cross sections from [25] to obtain the corresponding rate coefficient but not for the calculation of the EEDF. In Figure 14 we observe a decrease of the reduced atomic oxygen density when using Polak's cross sections and a slight reduction of the dissociation fraction, more important at higher $O_2$ content. The vibrational temperatures, $T_3$ and $T_{CO}$, are also impacted and increase for all conditions. Indeed, the main quenching mechanism for the $CO_2$ and CO vibrationally excited molecules occurs with atomic oxygen which becomes less important when the O/N decreases. From this analysis we can conclude that using Polak's cross sections leads to a decrease of O/N but further work is necessary to understand the validity of this cross section.

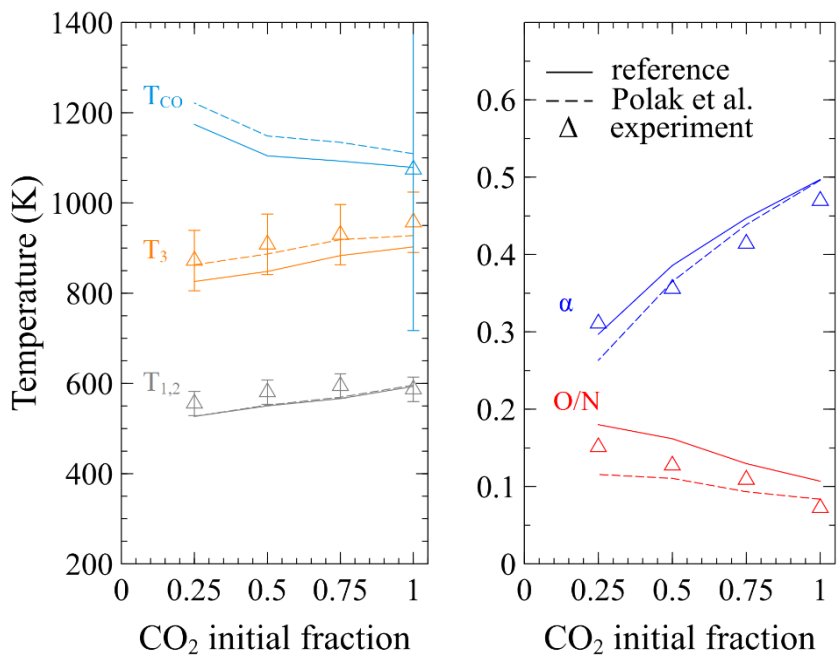

*Figure 14: Experimental values (Δ) and calculated values (line) of the common vibrational temperature of the $CO_2$ bending and symmetric modes $T_{1,2}$, the vibrational temperature of the asymmetric stretching mode $T_3$, the CO vibrational temperature $T_{CO}$, the dissociation fraction α and the reduced atomic oxygen density O/N, for a discharge ignited in different $CO_2$-$O_2$ mixtures, at 20 mA and 1 Torr, using our reference cross section (—) and the cross sections from Polak et al. [25] (– –) for $O_2$ dissociation.*

6) Dominant mechanisms

The model developed in this work allows a further understanding of the complex coupled plasma kinetics, providing estimations of excited species densities, reaction rates or electron properties but also the relative contributions of the different processes to the formation and loss of the species considered. Figure 15 depicts the contributions of the dominant creation and destruction mechanisms of $CO_2$ and CO for two extreme conditions of pressure (1 and 5 Torr), for pure $CO_2$ and a 50/50 $CO_2/O_2$ mixture, at 20mA. For each reaction we plot its relative importance for the creation (positive) or destruction (negative) of $CO_2$ (left panel) and CO (right panel). To facilitate this study, the results are shown for simulations where the vibrational kinetics are not included.

In our conditions, the main CO$_2$ dissociation mechanism is by electron impact at ~7 eV to create O($^1$D) and CO, and ground-state CO molecules are essentially created by dissociation through electron impact on CO$_2$ molecules. A dominant effect is the renewal of the gas (flow) controlling the loss of CO$_2$ and CO in this discharge.

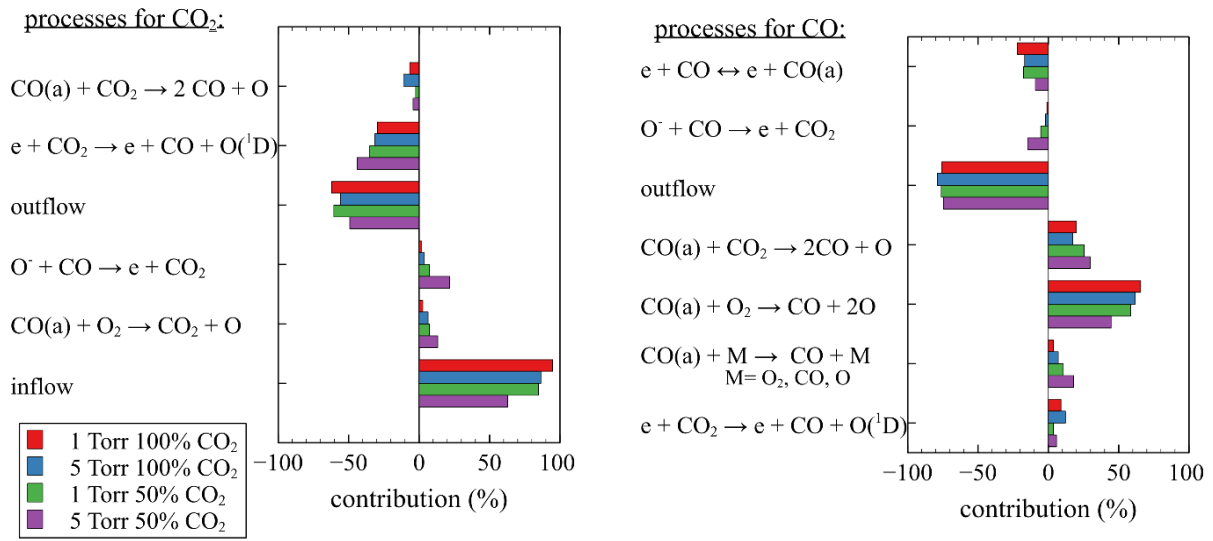

*Figure 15: Contribution of different processes for creation (+)/ destruction (-) of CO$_2$ (left panel) and CO (right panel) for a low pressure (1 Torr) and for a high pressure (5 Torr) conditions, at 20mA for pure CO$_2$ and a mixture of 50%CO$_2$ and 50% O$_2$.*

CO is also created from the quenching of the CO(a) state, mostly in collisions with CO and O, and to a lesser extend with CO$_2$ and O$_2$. However, CO(a) is obtained from the excitation of ground-state CO (via direct electron impact) which is one of the main processes of destruction of CO. Therefore, reactions involving only CO(a) and ground-state CO do not constitute true creation/destruction mechanisms of CO molecules, but only change the relative proportion of these two electronic levels. An effective creation mechanism of CO is the dissociation of CO$_2$ in collisions with CO(a), having a significant contribution to the production of CO. Finally, CO molecules are destroyed through the CO(a) state in the back reaction mechanism with O$_2$ giving back CO$_2$, and it also corresponds to one of the main CO$_2$ creation channels. In turn, CO$_2$ is mostly produced by renewal of the gas but also in the process CO(a) + O$_2$ → CO$_2$ + O, as mentioned above, contributing to more than 10% of CO$_2$ creation at 5 Torr, for the CO$_2$/O$_2$ mixture.

As could be expected, the addition of O$_2$ to the CO$_2$ plasma changes the relative contributions of the different processes. For instance, the two processes leading to CO$_2$ recombination, involve O$^-$ and O$_2$ species and gain importance when O$_2$ is added to the mixture, with respective contribution going from 4% to 22% and from 6% to 13%, at 5 Torr, whereas the contribution of the process leading to the dissociation of CO$_2$ by collision with CO(a) is decreased upon O$_2$ addition. Finally, O$^-$ ions influence the neutrals chemistry creating CO$_2$ back from CO in the recombination reaction. The effect of the negative O ions on the neutrals chemistry was already observed in [14] but at a higher current (50mA). Indeed, at low current, O$^-$ is mostly created by dissociative attachment with CO$_2$ and mainly destroyed in the reverse reaction. However, at high current or in our case, when O$_2$ is added to the mixture, the production of O$^-$ shifts towards dissociative attachment with O$_2$ and O$^-$ then reacts with CO producing CO$_2$.

The differences between the two cases of different presssure are not significant but the creation of CO by electron impact is enhanced at lower pressure. However, the quenching of CO(a) to CO is also more important at 1 Torr leading to a lower contribution of the CO(a) +$CO_2$ → 2CO + O reaction. Indeed, the main quenching of CO(a) occurs with CO and O and while the CO fraction remains almost constant with pressure, O/N is much higher at 1 Torr than 5 Torr (Fig. 9).

# V.   Conclusion

This work presents a model that includes the state-to-state kinetics of the first 72 low-lying levels of $CO_2$ corresponding to the vibrational levels with $v_1^{max}=2$ and $v_2^{max}=v_3^{max}=5$ and energies up to about 2 eV and the 10 first levels of CO as well as the chemical kinetics of $CO_2$ and dissociation products. It constitutes a step forward towards a more complete and thorough validation of $CO_2$ dissociation in low temperature plasmas. Indeed, we extended the model from Silva et al. [11, 17-19], previously validated for low pressure DC glow discharge in a $CO_2$ plasma, by including the CO vibrational kinetics (e-V, V-V and V-T), the deactivation of $CO_2$ vibrationally exited molecules in collisions with O, CO and $O_2$, and also the $CO_2$-CO V-V transfers, relevant in the context of $CO_2$ dissociation. For future studies, higher vibrational levels, up to the dissociation limit, should be included to better understand the underlying kinetics under a higher excitation regime. This should allow applying the model to plasma conditions targeted for $CO_2$ conversion on the industrial scale. This effort will involve the computation and validation of the rate coefficients involving highly vibrationally excited $CO_2$ molecules. However, first-order perturbation theories, like the SSH and SB approaches, while providing a good basis allowing for the description of $CO_2$ vibrations under low excitation regimes, cannot be used for the scaling of vibrational rates up to the dissociation limit. Different scaling procedures must be considered in future research.

The model was validated as a result of the good agreement between the calculated vibrational temperatures, O/N, E/N and dissociation fractions, and the corresponding experimental data measured in a DC glow discharge by in situ FTIR spectroscopy and actinometry. The reaction mechanism (validated set of reactions and corresponding rate coefficients) we propose predicts the quantities mentioned above for pressures between 0.4 and 5 Torr, discharge current of 20 and 40 mA and for different compositions ranging from 100% to only 25% of $CO_2$ in a $CO_2$-$O_2$ mixture.

The experimental trends associated with different pressures and mixtures were analysed. The experimental data show a lower conversion of $CO_2$ when $O_2$ is added to the plasma. The modelling study strongly suggests that this effect cannot be attributed to the quenching by O atoms of the vibrationally excited $CO_2$ but rather to enhanced back reactions involving the first electronically excited state of CO, CO(a), in combination with molecular oxygen or to a lesser extent with CO. Indeed, even though electronically excited states are often neglected in the study of plasma chemistry in $CO_2$ plasmas, they carry a significant amount of energy than can influence the heavy species chemistry under discharge conditions. When the CO and $O_2$ densities become large enough, an important contribution of back reaction mechanisms controlled by electronically excited CO have been demonstrated and the role of CO(a)+$O_2$ → $CO_2$ + O is especially relevant for $CO_2$-$O_2$ mixtures.

The similar thresholds for $CO_2$ dissociation through electron impact at ~7 eV and back reaction mechanisms controlled by electronically excited states of CO at ~6 eV suggest that effective separation of the dissociation products could enhance the $CO_2$ conversion efficiency. Future research should, therefore, concentrate on the development of separation procedure to isolate $O_2$ from the other dissociation products. Even though O atoms are not directly responsible for the reduced dissociation, their recombination at the wall to form $O_2$ is a key process [13, 31], and the use of membranes to extract O atoms from the plasma could thus enhance the conversion efficiency. Recent proposals for products separation include the use of silver membranes by Premathilake et al. and Wu et al. [97, 98], hollow fiber mixed-conductor membranes [99] and a new electrochemical membrane reactor presented by Goede and co-workers [7].

The choice of cross sections as well as the values of recombination probability of O at the walls are very important parameters which determine the atomic oxygen density in the discharge. The choice of the appropriate electron impact cross section for $O_2$ dissociation remains an open question, but the present work brings further insight into it.

The present results confirm the non-equilibrium nature of low-pressure $CO_2$ plasmas, with a characteristic temperature of CO, $T_{CO}$, well above the temperature of the asymmetric vibration mode, $T_3$, which in turn is above the vibrational temperatures of the other two modes, $T_{1,2}$, and the gas temperature, $T_g$. Moreover, this study also corroborates the importance of the vibrational transfer from CO to the asymmetric stretching mode of $CO_2$, of the quenching of vibrationally excited $CO_2$ and CO by O atoms and subsequent reduction of the $CO_2$-CO V-V, in an accurate description of the vibrational kinetics in $CO_2$ plasmas. For the current discharge configuration, $CO_2$ dissociation is driven by electron impact and vibrational excitation plays a negligible role in both the dissociation via the ladder climbing mechanism and in the back reaction mechanisms, due to the low excitation regime in the glow discharge. Nevertheless, vibrational kinetics has a significant influence in dissociation via the electron superelastic collisions with vibrationally excited CO and $CO_2$ molecules modifying the EEDF and leading to an increase of the electron impact dissociation rate coefficients and, accordingly, of the $CO_2$ dissociation.

Acknowledgments:
This work was partially supported by the European Union's Horizon 2020 research and innovation programme under grant agreement MSCA ITN 813393, and by Portuguese FCT-Fundação para a Ciência e a Tecnologia, under projects UIDB/50010/2020, UIDP/50010/2020 and PTDC/FIS-PLA/1616/2021. ASMC was funded by LabEx Plas@par receiving financial aid from the French National Research Agency (ANR) under project SYCAMORE, reference ANR-16-CE06-0005-01.